\documentstyle[fleqn,prb,aps,tighten,preprint]{revtex}

\def\br{{\bf r}}

\begin{document}
\newcommand{\eeq}{\end{equation}}
\newcommand{\intif}{\int_{-\infty}^{\infty}}
\newcommand{\sid}{\mbox{$\psi^{\dagger}$}}
\newcommand{\sib}{\mbox{$\overline{\psi}$}}
\newcommand{\il}{\int_{-\Lambda}^{\Lambda}}
\newcommand{\ie}{\int_{0}^{\Lambda}}
\newcommand{\iT}{\int_{0}^{2\pi}d\theta}
\newcommand{\iK}{\int_{\Lambda /K_F}^{\pi -\Lambda /K_F}}
\newcommand{\si}[2]{\mbox{$\psi_{#1}(#2)$}}
\newcommand{\beq}{\begin{equation}}
\draft
\widetext
\tolerance 50000
\title{
 Field Theory of the Fractional Quantum Hall Effect-I}
\author{G. Murthy$^{1}$  and
 R. Shankar$^{2}$
}
\address{$^{1}$ Department of Physics, Boston University, Boston MA 02215\\
and, Department of Physics and Astronomy, Johns Hopkins University, Baltimore
MD 21218\\
$^{2}$Sloane Physics Laboratories, Yale University, New Haven CT
06520\\
{\em (To appear in COMPOSITE FERMIONS, edited by Olle Heinonen.)}
}

\date{\today}
\maketitle
\widetext

\vspace*{-1.2truecm}

\begin{abstract}
\vspace*{-.5truecm}
\begin{center}
\parbox{14cm}{We present a     Chern-Simons (CS) theory for fractional quantum
Hall states,   in which
flux attachment is followed by an enlargement of   Hilbert space to include $n$
magnetoplasmon oscillators and
$n$  constraints, $n$ being the number of electrons. The  simplest
approximation
yields   correlated wavefunctions for fractions $\nu = p/(2ps+1)$.
For  $\nu =
1/(2s+1)$ we obtain      multi-quasihole wavefunctions  with
correct
gaussian and normalization factors,   operators
that create quaisholes, and  a  possible  Read operator. Our composite
fermion
and boson operators  attach flux {\em as well as the correlation hole
to
electrons, i.e., bind them to    zeros}. We perform a  further canonical
transformation which
decouples
the the particles and oscillators  in the long wavelength limit.  The     $n$
constraints  now act
on just the particles,  restricting them to  the  Lowest Landau
level (LLL).
 Their   kinetic energy vanishes In the noninteracting limit,   they
couple
to  the external  field through  a magnetic moment $e/2m$,
carry charge
$e^* = e/(2ps+1)$ and a dipole moment which alone survives when $\nu
=1/2$.
When a Coulombic interaction is turned on,  a
nonzero
$1/m^* \simeq  (2\nu )^{2/3} e^2l_0 /6$ arises. Density-density correlation
functions
and Hall conductivity,   additive over the particles and oscillators,
 are
calculated within an approximation for $\nu =1/2$. The formalism makes
precise
and unifies many prevalent notions   regarding composite fermions and
bosons,
quasiparticles, cyclotronic physics, $m$ and $m^*$. }

\end{center}
\end{abstract}

\narrowtext
\section{Introduction}

The spectacular discovery of the Fractional Quantum Hall Effect (FQHE)
by Tsui, St\"{o}rmer and Gossard\cite{fqhe-ex} has sparked a remarkable
amount of activity among theorists and
experimentalists\cite{review1,review2}.  The theoretical efforts fall
into two broad categories: those that seek trial wavefunctions that
capture the essential physics, for example references
\cite{laugh,halperin1,halperin2,haldane,girvin84,jain-cf,rezayi-read},
and those that attempt to construct a  theory that starts
with electrons in a magnetic field subject to Coulomb interactions and
try to work their way towards the experiments through a series of
approximations.  Chern-Simons (CS)\cite{CS} field theories play a
central role in the second
category\cite{zhk,kane,lopez,kalmeyer,hlr,brad}. The present
article, also of this genre, is an elaboration of a recent
letter\cite{prl}. It relies on its predecessors, extends them, unifies
many miscellaneous ideas that have been in the air and exposes the
underlying physics in a particularly transparent way.

We follow the standard notation in which the electrons, $n$ in
number,  have a
bare (band) mass $m$, charge $-e$,
 a complex coordinate $z=x+iy$, move in unit area in  the $x-y$
plane, and are
in a field ${\bf B} = -\hat{z} B$ pointing  down the $z-$ axis. We
set $\hbar =
c = 1$. The inverse  filling fraction
\beq
\nu^{-1} = {eB\over 2\pi n}
\eeq
gives the number of flux quanta per electron.
In the cases of interest, $\nu <1$, it will be assumed all spins are
polarized.

Let us first neglect  interactions.  All the electrons will populate the Lowest
Landau Level
(LLL). The ground state will be macroscopically degenerate and the
first
nonzero energy
 excitation will be at the cyclotron frequency
\beq
\omega_c = {eB\over m}.
\eeq
The effective mass of the particles will be  infinite:
\beq
{1 \over m^*} = 0
\eeq
reflecting the quenching of kinetic energy.

 Let us now turn on a Coulomb interaction, weak in the sense
\beq
e^2/l_0 << \omega_c
\eeq
where the cyclotron length
\beq l_0 =
(eB)^{-1/2}
\eeq
is a measure of the typical interparticle spacing and
 the size of the orbits, and where the dielectric constant $\varepsilon$ has
been suppressed.  One may reasonably expect that in the limit $m\to 0$ all
higher Landau levels and any reference to $m$ will disappear from the
picture; that the degenerate manifold of the LLL states will broaden
out to a width controlled by the effective mass $ 1/m^* \simeq
e^2l_0$, of Coulombic origin.  Thus consider   Laughlin's
wavefunctions\cite{laugh} at $\nu = 1/3$ for the ground state, and  state with
a quasihole at the origin:
\begin{eqnarray}
\Psi_ {1/ 3}  &= & \prod_{i<j} (z_i - z_j) ^{3}\exp{-\sum_i
|z_i|^2/(4l_{0}^{2})}\\
 \Psi_{1/3, qh}  &= & \prod_k  z_k\prod_{i<j} (z_i - z_j)
^{3}\exp{-\sum_i
|z_i|^2/(4l_{0}^{2})}.\label{eq-laugh}
\end{eqnarray}
Note  that only   $l_0$ enters the wavefunctions and  $m$ does not. This is
true also of the subsequent functions written down by
Haldane\cite{haldane}, Halperin\cite{halperin1,halperin2},
Jain\cite{jain-cf} and Rezayi-Read\cite{rezayi-read}.

{}From these
wavefunctions one has gleaned many of the basic features of the FQHE such as
the incompressibility\cite{laugh}, fractional charge of the
quasiparticles\cite{laugh,halperin1,halperin2,jain-cf}, and their fractional
statistics\cite{halperin2,arovas}.
These successes notwithstanding, there are many reasons to pursue the
hamiltonian  approach in parallel. If it  can
yield these wavefunctions in some approximation, the nature of that
approximation will be physically revealing. One can
calculate other  wavefunctions and  correlation functions at general
$(q,\omega )$, one can couple the system to impurities to study the effects of
disorder, or
to a confining potential to study the physics at the edge. The
underlying  theory will tell us how the charge of the excitations gets
renormalized from $e$ in the case of free electrons to $e^* = e/(2p
s+1)$ for the principal fractions $p/(2ps+1)$, and in particular how
it vanishes for the case of $\nu =1/2$,  becoming
dipolar as anticipted by Read\cite{read2}. It should also tell us who carries
the Hall
current if the quasiparticle (the composite fermion) is neutral.  It
should shed light on the order parameter with algebraic order proposed
by Girvin and MacDonald \cite{gm} and studied by ZHK\cite{zhk}, and
the order parameter with long-range order proposed by
Read\cite{read1}.

The  theory must also address the question of the bare and
renormalized masses in some detail.
The notion that low energy physics can be described with no reference
to $m$ or higher Landau levels is plagued with subtleties as
emphasized by Girvin, MacDonald and Platzman
(GMP)\cite{GMP}. First, the current operator depends explicitly on $m$
while the Hall conductance $\sigma_{xy}$ does not.  For $m$ to drop
out, one must necessarily invoke virtual transitions to the
next Landau level\cite{LLLcurrents} with matrix elements of order
$m$. Next, one must understand  how $m$ gives way to $m^*$  in the rest of
the low energy physics and how the latter is to
be calculated in terms of the interactions. Even if one begins with a
phenomenological $m^*$, there are places where $m$ enters. For
example, the cyclotron pole is determined by $m$ according to  Kohn's
theorem\cite{kohn}. In the Fermi-liquid-like state of $\nu =1/2$,
Simon and Halperin\cite{simon} argue that it is necessary to bring $m$
back by invoking a suitable Landau parameter $f_1$ and the relation

\beq
{1\over m} = {1 \over
m^*} + {f_1 \over 2\pi},
\eeq
a scheme that was briefly alluded to in HLR\cite{hlr}.
Now one may argue that the Kohn mode is
part of the high energy physics and an effective low energy theory is
not obliged to explain it. But as Simon, Stern and Halperin (SSH)
\cite{SSH} point out, $m$ enters the low energy physics as
well. Consider the zero point energy of $eB/2m$ per electron, which one
may  ignore as a constant. Suppose we impose on the system a
small slowly varying field $\delta B (r)$. This amounts to imposing a
potential $e\delta B(r)/2m$, the response to which is surely $m$
dependent. Also, if for any reason the density becomes nonuniform,
there will be a current due to incomplete cancellation of cyclotron
currents \cite{GMP}:
\beq
{\bf j}_{mag} = {e \over 2m} \hat{\bf z} \times \nabla n .
\eeq
Both these effects imply that the charge-current correlator $K_{01} = K_{10}$
has a piece $ (iqe/2m)K_{00}$ as $q\to 0$.
SSH show that these effects can be reproduced by appending to each
particle a magnetic moment $e/2m$. It is desirable to have this
occur naturally within a  computational scheme.

The variant  of  CS theory we propose here solves many of these
problems.

To set the context, let us begin by recalling the basic idea behind CS
theories, using as a concrete example $\nu ={1/ 3}$ and the goal of
getting Laughlin's wavefunction $\Psi_{1/ 3}$. Clearly, the more we
know of the answer the less we have to calculate. The usual bosonic CS
theory of Zhang, Hansson and Kivelson (ZHK) \cite{zhk} begins by
writing the wavefunction for the state as
\beq
\Psi_{1/ 3} ({\bf r}_i) = \left[ phase \ of \ \Psi_{1/ 3}\right]
\Psi_{CS}({\bf
r}_i)  \label{pref}
\eeq
and then solving for $\Psi_{CS}$ which describes CS bosons.  At mean field
level the bosons condense into the zero momentum state, giving
$\Psi_{CS}=1$ and we obtain just the phase of $\Psi_{1/ 3}$. Each
particle sees three flux tubes in the others, but does not avoid them
with the triple zeros, a correlation essential to the success of
Laughlin's wavefunction. These zeros, which imply correlation holes,
arise upon going beyond mean-field and including fluctuations
\cite{kane}. This is also true for fermionic CS theories of Lopez and
Fradkin\cite{lopez}, Kalmeyer and Zhang\cite{kalmeyer}, Halperin, Lee
and Read (HLR)\cite{hlr} and Kwon, Marston and Houghton\cite{brad},
though wave functions have not been explicitly worked out this way for
the case of $\nu =1/2$.  For the same  reason, the composite bosons and
fermions all carry the full charge of the electrons at mean-field
level and the fractional charges of the quasiparticles are again to be
found in the fluctuations.  An exception is the theory of
Rajaraman and Sondhi\cite{rajaraman}.  These authors include all of
$\Psi_{1/3}$ as a prefactor in Eqn.(\ref{pref}) so that at mean field
level they obtain all of $\Psi_{1/ 3}$.  On the other hand, the
fluctuations about mean field are described by a complex CS field
whose real and imaginary parts come from the phase and modulus in the
prefactor.\footnote{A CS theory with  complex electric and magnetic fields  was
briefly  discussed  up by Girvin in the closing article in Reference
(\ref{gir}).
} To go  beyond mean-field theory using diagrams one must face the fact that
the hamiltonian is not hermitian. However the
   formalism has already been applied  successfully to  questions involving
solitons\cite{rajaraman}.

The method we propose manages  essentially to append as a prefactor
both the
modulus and phase of $\Psi_{1/ 3}$ (and its generalization to
fractions of the
form $p/(2ps+1)$) {\em without the use of complex vector potentials.
} The
correlation holes give the right charge $e^*$ to the quasiparticles.
The
theory has two sectors, and with one (desirable) exception,  $m$
appears in
one and  $m^*$ in the other.

Setting aside details for later, here is how this is done. Recall (or
go and read, if you are too young) the work of Bohm and
Pines\cite{bohm-pines} on the electron gas in three
dimensions. Besides its fermionic quasiparticles and particle-hole
excitations, the gas has plasmons in the spectrum.  These arise as
poles in the density-density correlations. The poles have a negligible
imaginary part (i.e., the plasmons are very long-lived) till the electron-hole
continuum
runs into the plasmon pole at large enough $q$. Rather than treat the
small $q$ plasmons as  composites built out of the original degrees of
freedom, Bohm and Pines elevate them to elementary particles with their
own set of states and operators in an enlarged Hilbert space.  The
plasmons couple to the electrons in way that reproduces the pole and
residue in the density-density correlations. Having introduced the
plasmons as elementary objects, one must ensure that the particles do
not duplicate them.  To keep the problem unchanged in the enlarged
Hilbert space, only states obeying a set of constraints (one for each
${\bf q}$ at which a plasmon was introduced) are deemed physical. The
constraints freeze all fermionic collective degrees of freedom that
were assigned to plasmons. Placing the plasma oscillators in their
ground states gives, {\em upon projection to the physical states}, a
correlation factor to the fermionic wavefunction, which is simply the
exponential of the Coulomb potential energy associated with each
configuration. If the principal coupling between the fermions and
plasmons is removed by a canonical transformation, it is found that
the fermion mass gets renormalized upwards while the plasmon pole is
fixed at the classical value as $q\to 0$. The number of plasmon modes
introduced is treated as a variational parameter and is a small
fraction of the number of fermions.

We adapt this scheme to our problem with the magnetoplasmons playing
the role of plasmons. {\em For a variety of reasons we pick $n_0$, the
number of oscillators, to equal $n$, the number of particles. }
Putting the magnetoplasmons in their ground states gives the
correlation zeros of the trial wavefunctions. Our field operators for composite
bosons and fermions
create the electron as well as its correlation hole, the latter being
described by collective coordinates. When we decouple the oscillators
from the particles, $1/m^*$ gets renormalized to zero in the
noninteracting limit and picks up a finite value when interactions are
turned on. The zeroth order approximation to $1/m^*$ is readily
evaluated.  The decoupling, performed in the  long-distance limit and the RPA,
also changes the charge of the quasiparticles from $e$
to $e^*$.  Upon decoupling, the $n$ constraints act on the particles
alone,  the reduction in degrees of freedom being exactly what it takes
to get LLL physics.  This is confirmed by the fact that the density (limited to
the sector in which the oscillators are in their ground states) now obeys the
magnetic translation
algebra of GMP\cite{GMP}. All dependence on $m$ is contained in the
oscillator sector with one exception: the particles couple to the
external magnetic field with a moment $e/2m$, the need for which was
discovered  by SSH\cite{SSH}. We find that in the
clean limit, all the Hall current is carried by the oscillators,
which takes care of the nagging question of who carries the Hall
current when the composite fermions become neutral at $\nu =1/2$.  Our
formalism gives a precise meaning to the very useful concept of the
composite fermion .
All this will be explained as we go along.

The plan of the paper is as follows: In Section II we recall the key
ideas behind flux attachment in both the wavefunctions and CS
approaches. Section III describes our CS representation which results
from enlarging the Hilbert space and doing a canonical transformation
to reach what we call the middle representation (MR). In Section IV we
derive correlated ground state wavefunctions for $\nu = 1/(2m+1), 1/2$
and the Jain series $p/(2p+1).$ (The case $p/(2ps +1)$ calls for a
straighforward generalization and we do not discuss it in any detail.)
For the first case, described by a bosonic CS theory, we display
operators that create quasielectrons and quasiholes and a candidate
for the Read operator\cite{read1}.  We show that our composite
boson and fermion operators create particles in which an electron is
bound to zeroes, and not just flux quanta. In Section V we perform the
canonical transformation to the final representation (FR) in which the
particles and oscillators are decoupled in a long-distance
approximation. We exhibit the substantial mass and charge
renormalization. We pay special attention the gapless case $\nu =1/2$
which arises in the limit $p\to \infty$ and compare it to the work of
HLR.  Discussions and conclusions follow in Section VI.

\section{Flux attachment and Chern-Simons Theories}

In two space dimensions one has the remarkable freedom to alter
particle statistics by attaching point flux tubes\cite{leinaas,wil}.
In particular, a fermion carrying an odd/even number of flux quanta
will behave like a boson/fermion.  This fact is exploited as
follows. Let us consider $\nu =1/3$. In the noninteracting limit, the
ground state is macroscopically degenerate. Trading the electrons for
composite bosons (CBs), each carrying three quanta of statistical flux
opposed to the external field, one ends up {\em on the average} with
composite bosons in zero field. Alternatively, trading them for
composite fermions (CFs) carrying two quanta (opposing the external
field) each, one obtains CFs with one quantum each on the average,
i.e., enough to fill exactly one Landau level. In either case we have
a nondegenerate starting point for including the effect of
interactions. The notion of dealing with the flux on the average to
get a simple starting point was first exploited by
Laughlin\cite{fetter} in studying anyon superconductivity. Of course
one must deal with the (possibly singular) fluctuations to complete
the analysis of the problem.

Let us see how flux attachment  is implemented in building wavefunctions and
CS field theories.

 It was noted long ago by Halperin\cite{halperin1} that the $3n$ zeros
of Laughlin's wavefunction are all tied to the location of the
particles, while only $n$ are required to do so by Fermi
statistics. The electron and the triple zero are attracted to each
other by the Coulomb force, and behave like a neutral boson, which
condenses in the Laughlin state, as emphasized by Read\cite{read1}.  Thus one
can write
\begin{equation}
\Psi_{1/3} =  \prod_{i<j} (z_i - z_j) ^{3}\exp{\left[ -\sum_i
|z_i|^2/(4l_{0}^{2})\right] } \equiv   \chi^{3}_{1}  \cdot 1,
\end{equation}
where $\chi^{3}_{1}$ attaches three flux quanta  to the CB whose condensed
wavefunction is simply the factor of unity.

On the other hand for fractions of the form $p/(2ps+1)$ one does not
have such explicit wavefunctions. Here is where Jain's notion of
composite fermion (CF) \cite{jain-cf} comes to the rescue.  First he
notes that we can just as well write
\begin{equation}
\Psi_{1/3}= \prod_{i<j} (z_i - z_j) ^{3}\exp{-\sum_i
|z_i|^2/(4l_{0}^{2})}
 = \chi^{2}_{1} \cdot \chi_1.
\end{equation}
The picture now is that the electrons have been traded for CF's which carry two
flux quanta opposed to the external field, and fill exactly one Landau level in
the weakened mean-field.
 The first factor of $ \chi^{2}_{1}$ attaches two flux quanta and the
last stands for the filled Landau Level of the CF.  Now if $\nu =
p/(2ps +1)$, with $p>1$, bosons in zero field are not attainable, but
Jain's idea composite fermion (CF) idea is still applicable. By attaching $2s$
units of flux
opposed to the external one, the electron problem gets mapped on
average to that of composite fermions which see
\begin{equation}
{1 \over \nu^*} = { 1 \over \nu} - 2s =1/p
\end{equation}
units of flux each, i.e., fill $p$ Landau levels. For example if $\nu
=2/5$, we
have $p=2, s=1$ and the Jain wavefunction is
\begin{equation}
\Psi_{2/5} =  {\cal P} \chi^{2}_{1} \chi_2
\end{equation}
where $\chi_2$ describes two filled Landau levels and ${\cal P}$
projects the wavefunction to the LLL. (The magnetic lengths appearing
in the gaussians in $\chi_1$ and $\chi_2$ must be chosen to reflect
the magnetic fields associated with filling one and two levels
respectively.) When $\nu =1/2$, (the $s=1$, $p \to \infty $ limit) the
external and attached fluxes cancel and we end up with
\begin{equation}
\Psi_{1/2} =  {\cal P} \chi^{2}_{1} \ \ |FS\rangle
\end{equation}
where $|FS\rangle$ stands for the Fermi Sea. This function was
proposed
independently and studied
in detail by Rezayi and Read (RR) \cite{rezayi-read}. They showed that  a gap
is not needed for the success of the idea.

{\em It is important to note that there is no unique notion of flux
attachment in the wavefunction approach.}  For example, instead of
using the factor $\chi_1$ to attach a unit of flux one could use a
factor that did not contain the gaussian, or even the zero when the
particles approached each other, by the making the replacement
\begin{equation}
(z_i - z_j)  \to {(z_i - z_j) \over |z_i -z_j|} .
\end{equation}
If the latter choice had been made for $\nu =1/3$ one would have lost
two of the three zeros that produced the excellent correlations as the
particles approached each other. Of course, such a wavefunction would
have been a much poorer approximation to the true ground state than
one where the correlation zeros were present. There is no reason to
make the poor choice in the quest for wave functions, but as we shall
now see, the choices are rather limited in the  CS
approach to which we now turn.

In contrast to the, flux attachment in CS field
 theories is a lot more restrictive, and is usually
carried out as follows. In
 first quantization one introduces a wavefunction for the CS particles
 in terms of the electronic wavefunction as follows\cite{leinaas}:
\begin{equation}
\Psi_e =
\prod_{i<j} {(z_i - z_j) ^{l}\over |z_i
-z_j|^{l}}\Psi_{CS}.\label{eq-phase}
\end{equation}
where $l$ is the number of flux quanta to be attached.  The CS
particles are to be quantized as fermions/bosons, for $l=
$even/odd since the prefactor produces a factor $(-1)^l$ under
particle
exchange. The prefactor   introduces a gauge field ${\bf a}$ in the
Schr\"odinger equation for $\Psi_{CS}$. The field obeys
\begin{equation}
{\nabla \times
{\bf a}\over 2 \pi l} = \sum_{i}^{n} \delta^2({\bf r} - {\bf r}_i)
.\label{eq-const1}
\end{equation}
In second quantization the CS transformation is represented by the
operator
relation
\begin{equation}
\psi_e ({\bf r}) = \exp \left[ {i l \int {{\hat
{\bf z}}\times(\br-\br')\over|\br-\br'|^2}\rho ({\bf r'}) d^2r'}\right]
\psi_{CS} ({\bf r'})
\end{equation}
where $\psi_e$ and $\psi_{CS}$ are the field operators that annihilate
the electron and the CS composite particle respectively. (They are not
to be confused with the wavefunctions $\Psi $). Also,
\begin{equation}
\rho ({\bf r}) = \psi_{e}^{\dag}({\bf r}) \psi_{e}^{}({\bf r}) =
\psi_{CS}^{\dag}({\bf r}) \psi_{CS}^{}({\bf r})
\end{equation}
is the density. The hamiltonian(density), before turning on
electron-electron interactions (which will be added later), is
\begin{equation}
H_{CS} = \psi^{\dag}_{CS}
{|(-i\nabla +e {\bf A}
+{\bf a})|^2 \over 2m}\psi_{CS}
\end{equation}
where ${\bf A}$ is the external vector potential, $m$ is the bare
mass and

\begin{equation}
{\nabla \times
{\bf a}\over 2 \pi l} = {\psi^{\dag} }\psi .\label{eq-const2}
\end{equation}

Since ${\bf A}$ describes a magnetic field pointing down the $z$-axis,
and the CS field has its flux of $l$ quanta per particle pointing up,
they can be played off against each other.  There are two kinds of
$\nu$ for which things are particularly simple. If $\nu = 1/(2s +1)$,
by attaching $2s+1$ quanta per particle, we can trade the electrons
for bosons that see zero net field {\em on the average}. If $\nu =
1/2s$, by adding $2s$ quanta per particle we can trade the electrons
for fermions in zero average field. Hereafter we use the convention
that the average values have been removed from ${\bf a } $ and
$\psi^{\dag} \psi = \rho $ (the density) both of which henceforth
refer to normal-ordered quantities. (The normal ordering will occasionally be
 made explicit).  The bosons  condense in zero field
and the fermions form a Fermi sea (FS), both of which are nice
nondegenerate starting points explored by Zhang {\em et al}\cite{zhk,kane}
(bosonic) and Kalmeyer and Zhang\cite{kalmeyer}, HLR\cite{hlr}, Kwon {\em et
al}\cite{brad}
(fermionic). At the mean-field level the wavefunctions one obtains are
\begin{equation}
\Psi_{1/l} = \prod_{i<j} {(z_i - z_j) ^{l}\over |z_i
-z_j|^{l}}\Psi_{CS}
\end{equation}
where $\Psi_{CS}$ is the Fermi Sea (FS) for $\nu =1/2$ and $\Psi_{CS}
= 1$ for $\nu =1/(2s +1)$. The multiple  zeros expected in the
correlated  wavefunctions are absent.  To obtain these one needs to
analyze
fluctuations about the mean field\cite{kane}.

When $\nu = p/(2ps+1)$, by attaching $2s$ opposing  flux quanta per
electron,
we can cancel enough of the external flux so that what remains,
$e{\bf A}^* $,
is enough to fill exactly $p$ Landau levels :
\beq
eB^* = {eB\over 2ps+1} \label{bstar}.
\eeq

These are the fractions considered by Jain\cite{jain-cf}. Indeed
Jain's case reduces to the two previous ones when $p=1$ and $p=\infty$
respectively. The corresponding field theory was analyzed by Fradkin
and Lopez\cite{lopez} who computed some response functions and also
showed that fluctuations produced the modulus of the Laughlin function   in the
limit
of  particle separation large compared to $l_0$.

Let us summarize: In the quest for wavefunctions we can build in not
just the phase, but all of $\prod (z_i -z_j)^l$, or even the
ubiquitous gaussian factors into the process of flux attachment.  In contrast,
the standard  way of attaching flux in CS
theories leads only to the phase of the zeros at mean-field level,
while the zeros themselves emerge from the
fluctuations\cite{kane,lopez}.  It is possible to obtain the zeros in
a field theory at mean-field level: however, doing so leads to a
complex vector potential and a non-hermitian hamiltonian, as pointed
out by Rajaraman and Sondhi who analyzed this possibility in some
depth\cite{rajaraman}.
Let us now turn to our version of the CS theory and see how correlated
wave functions are obtained.

\section{Our CS theory}
Let us recall the CS hamiltonian (density) before introducing any
interactions:
\begin{equation}
H_{CS} = \psi^{\dag}_{CS}
{|(-i\nabla + e{\bf A}^*+ {\bf a})|^2 \over
2m}\psi_{CS}.\label{prettyboy}
\end{equation}
Here ${\bf a}$ refers to the fluctuations about the mean (which  has
been used to reduce ${\bf A}$ down to ${\bf A}^*$), and obeys
\begin{equation}
{\nabla \times
{\bf a}\over 2 \pi l} = :{\psi^{\dag} }\psi  : .\label{eq-const3}
\end{equation}
To make explicit the fact that ${\bf a}$ is really a dependent field,
let us rewrite this as

\beq
H = {1 \over 2m} \sid_{CS} (-i\nabla + e{\bf
A}^*+ (\nabla \times)^{-1} 2\pi l\rho )^2\psi_{CS} \label{curl-inv}
\end{equation}
where the inverse curl is uniquely defined by its Fourier transform if
we demand that  the answer be transverse. If you find this form of $H$
forbidding, it is our intention; Eqn.(\ref{prettyboy}) hides the
complexity of the CS hamiltonian.

We are now ready to enlarge the Hilbert space. Consider a disk (in momentum
space) of
radius $Q=k_F$, the Fermi momentum of the electrons, which contains as
many points as there are electrons.  For each ${\bf q}$ in this disc,
we associate a canonical pair of  fields $a({\bf q}),P({\bf q})$
\beq
\left[ a({\bf q}), P({\bf q}')\right] =
(2\pi )^2 \delta^2({\bf q}+{\bf q}') .
\eeq

These define a pair of longitudinal and transverse vector fields
\begin{equation}
{\bf P (q)} = i\hat{{\bf q}} \  P({\bf q}) \ \ \ \ \ \ \ \ {\bf a
(q)} = -i
\hat{{\bf z}}
\times \hat{{\bf
q} }\  a({\bf q}).\label{pq}
\end{equation}

The Hamiltonian density of Eqn. (\ref{curl-inv}) is completely
equivalent to
\beq
H = {1 \over 2m} \sid_{CS} (-i\nabla + e{\bf A}^* + {\bf a}+  (\nabla
\times)^{-1}
2\pi
l\rho
)^2\psi_{CS} \label{extendedH}
\end{equation}
provided  we restrict our selves to  states in the larger space obeying
\beq
a({\bf q}) |physical \rangle =0 \ \ \ \ \ \ \ \ q<Q.
\eeq

In other words, $\left[ H, a\right] =0$ allows us to find simultaneous
eigenstates of $H$ and $a$, and the eigenstates with $a=0$ solve the
original problem.

Let us note for future use how we are to extract wavefunctions for
the CS
particles from any solution in the bigger Hilbert space. Let
$\Psi_{CS} (x) $
denote the CS wavefunction. In the larger space we have, in obvious schematic
notation,  the following resolution of the identity
\beq
 I = \int dx da |x\ a\rangle \ \langle a\ x|
\eeq
where $x$ denotes all the particle coordinates and $a$ stands for
$a(q), [
0<q\le Q]$. The projection operator to the  physical sector is
\beq
\wp = \int dx da |xa\rangle \ \langle ax| {\delta(a)\over \delta(0)}.
\eeq
This means that if $\Psi(x,a)$ is a generic wavefunction, then the
projected
version is
\beq
\Psi_{\wp } (x,a) = {\delta(a)\over \delta(0)} \Psi (x,a) =
{\delta(a)\over
\delta(0)} \Psi (x,0). \label{physical}
\eeq
Indeed one can show that  every physical vector must have such a
form.

Let $\Psi_{\wp } (x,a)$ obey the eigenvalue equation (in first
quantization)
\beq
H(x,p,a) \ \Psi_{\wp } (x,a) = E \ \Psi_{\wp } (x,a)
\eeq
where $H(x,p,a)$ is the hamiltonian with $a$ added on as in
Eqn.(\ref{extendedH}).
Such eigenfunctions will exist since $H$ commutes with the
constraint. Feeding
in
Eqn.(\ref{physical}) for the wavefunction, moving  the delta function
through
$H$ to its left using the commutativity, and canceling it from both
sides,  we
obtain
\beq
H(x,p,0) \ \Psi (x,0) = E \ \Psi (x,0)
\eeq
establishing that
\beq
\Psi_{CS}(x) = \Psi (x,0).
\eeq
In other words $\Psi_{CS}$ is embedded within $ \Psi_{\wp } (x,a) $
as follows:
\beq
\Psi_{\wp } (x,a) = {\delta(a)\over \delta(0)} \Psi_{CS} (x).
\label{psiformula}
\eeq

Let us now return to the Hamiltonian density of Eqn. (\ref{extendedH}) which
contains the very nasty inverse curl. We deal with it through  the following
unitary transformation
\beq
U = \exp [\sum_{q}^{Q} i
P(-q){2\pi l\over q} \rho (q) ] .
\eeq
 In the above and what follows $q$ sums stand for integrals
\beq
\sum_q = \int {d^2q\over 4\pi^2}
\eeq
and the  vector nature of ${\bf q}$ is often suppressed.  Under the action of
$U$
\begin{eqnarray}
  U^{\dag} a (q) U &=  &  a(q)  - {2\pi l \rho  (q) \over
q}\\
\psi_{CS} (x) &=& \psi_{CP} (x) \exp [\sum_{q}^{Q} i P(-q){2\pi
l\over q}
e^{-iqx} ] \label{cstocb}\\
\!\!\!\!\!\! \psi_{CS}^{\dag}(x) (- i\nabla ) \psi_{CS}(x) &= &
\psi_{CP}^{\dag}(x) ( -i\nabla  + 2\pi l {\bf P}(x)) \
\psi_{CP}(x)\\
H & = & {1 \over 2m} \psi_{CP}^{\dag} (-i\nabla + e{\bf A}^* + {\bf
a}+  2\pi l
{\bf
P} + \delta {\bf a})^2\psi_{CP} \label{eq-hfree}\\
0&=& (a -  {2\pi l \rho \over q})|physical \ \rangle \ \ \ \ \ \
0<q\le Q
\end{eqnarray}
where $\delta {\bf a}$ refers to the {\em dependent} short range
vector potential $(\nabla \times)^{-1} 2\pi l\rho $ for $q>Q$ that did
not get cancelled by the unitary transformation, and $\psi_{CP}$
refers to the composite particle (boson or fermion) field.  The above
description of the problem, after the canonical transformation by $U$,
will be referred to as the {\em middle representation (MR)}. We argue
that $\Psi_{CP}$ better describes the composite bosons and fermions
alluded to in the literature than does $\Psi_{CS}$. Whereas the latter
are associated with particles carrying just flux tubes, the former are
associated with particles that carry flux tubes {\em and the
correlation holes}, i.e., describe electrons bound to zeros, as we shall see.

Suppose we have a physical  wavefunction $\Psi_{\wp}^{MR}(x,a)$ in
the  MR:
\beq
\Psi_{\wp }^{MR} (x,a) = {\delta (a - {2\pi l\rho \over q})\over
\delta (0)}
\Psi^{MR} (x,a)
\eeq

What is the corresponding CS wavefunction? Given that $U$ implements
a
translation of $a(q)$ by $-{2\pi l\rho \over q}$, it follows that
\beq
\Psi_{\wp } (x,a) = \Psi_{\wp }^{MR} (x,a +{2\pi l\rho \over q}
)={\delta
(a)\over \delta (0)}\Psi^{MR} (x,a + {2\pi l\rho \over q}) = {\delta
(a)\over
\delta (0)}\Psi^{MR} (x,{2\pi l\rho \over q}) \label{psiform}
\eeq
from which follows, upon invoking Eqn.(\ref{psiformula}),  the
invaluable
result
\beq
\Psi_{CS}(x) = \Psi^{MR} (x,{2\pi l\rho \over
q}).\label{psiformulaMR}
\eeq

The hamiltonian in Eqn.(\ref{eq-hfree})
\beq
H  =  {1 \over 2m} \psi_{CP}^{\dag} (-i\nabla + e{\bf A}^* +{\bf a}+
2\pi l
{\bf
P} + \delta {\bf a})^2\psi_{CP}
\eeq
has a local gauge invariance under
time-independent
transformations:
\begin{eqnarray}
\psi_{CP} &\to & e^{2\pi il\Lambda} \psi_{CP}\\
{\bf P} &\to   &{\bf P } - {\bf \nabla }\Lambda
\end{eqnarray}
where $\Lambda$ has only Fourier modes with $q\le Q$.  The constraint
merely
states that physical states
must be
singlets of the generator of these transformations:
\begin{equation}
({\nabla \times {\bf a}\over 2 \pi l} - :{\psi^{\dag} }\psi :)|
physical \
\rangle =0.\label{eq-conststates1}
\end{equation}
We may also write the above as
\begin{equation}
({q a\over 2 \pi l} - :\rho (q) : )| physical \
\rangle =0 \ \ \ \ 0 < q \le Q.\label{eq-conststates2}
\end{equation}

Had we chosen $Q$ to be infinite, $\delta {\bf a}$ would have vanished
and we would have entered what is called the Weyl or $a_0 =0$
gauge. In the Bohm-Pines case, $Q$ is very small compared to $k_F$ and
chosen as a variational parameter\cite{bohm-pines}. In our problem the
choice $Q=k_F$ recommends itself over all others repeatedly, as we
shall see.  The gauge we use is a nonstandard one, tailor made for
this problem. Sometimes one is asked how a choice of gauge could
matter, given that the physics is gauge invariant.  This misses the
point, which is that one can use the gauge freedom to highlight some
specific feature of the theory. Thus in Yang-Mills theories, the
Coulomb and unitary gauges display the physical degrees of freedom,
the Feynman-Lorentz gauges the Lorentz invariance, the $R_{\xi}$ gauge
the renormalizability etc.

We now explore our hamiltonian Eqn.(\ref{eq-hfree}), expanding  it as
follows (and dropping the subscript CP):
\begin{eqnarray}
H &=& {1 \over 2m}|(-i\nabla  + e{\bf
A}^* + \ {\bf a} +  2\pi
l{\bf P}+\delta {\bf a})\psi |^2
\label{eq-hamo}\\
& =&  {1 \over 2m} |(-i\nabla +e{\bf A}^* )\psi|^2 +{n \over 2m}(a^2+
4\pi^2l^2
P^2)
\nonumber \\
& +&  ({\bf a} +2\pi l{\bf P}) \cdot {1 \over 2m} \psi^{\dag}(-i
\stackrel{\leftrightarrow}{\nabla} +e{\bf A}^*)\psi \nonumber \\
 &+& { :\psi^{\dag}\psi :\over 2m}
({\bf a} +2 \pi l {\bf { P} })^2 \nonumber \\
&+&  {{\bf \delta a} \over 2m}\cdot \psi^{\dag}(-i
\stackrel{\leftrightarrow}{\nabla} + e{\bf A}^*)\psi   +
 { \psi^{\dag}\psi \over 2m}
(2({\bf a} +2 \pi l {\bf { P} })+ \delta {\bf a})\cdot \delta {\bf
a}\\
&\equiv &H_0 + H_I + H_{II} +H_{sr}\label{eq-ham}
\end{eqnarray}
in obvious notation, $H_{sr}$ being the terms associated with the
nondynamical
short-range
gauge field $\delta {\bf a}$. Note that we are yet to add
interactions.

\section{Wavefunctions for ground states and excitations}

In this section we will show how the above hamiltonian leads to some
well known correlated wavefunctions upon making the simplest
approximation. Let us first consider the fractions $\nu = 1/(2s +1)$
and $1/2s$ for which ${\bf A^*}=0$.  After this we will turn to Jain's
principal fractions $p/(2ps +1)$.  We will focus on the case $s=1$
since larger values of $s$ do not tell us anything new.
\subsection{$\nu =1/3$}
For this case we trade the fermions for bosons by attaching $l=3$ flux
quanta.  {\em Let us focus our attention on just $H_0$ in
Eqn.(\ref{eq-ham})}. It reads

\beq H_0= {1 \over 2m} |(-i\nabla
)\psi|^2 +{n \over 2m}(a^2+ (6\pi P)^2 ).
\label{hnot}
\eeq

There are no cross terms between ${\bf a}$ and ${\bf P}$ since they
are transverse and longitudinal respectively.  Given that $(a \ , P)$
are canonically conjugate, we see that the second term describes
harmonic oscillators of cyclotron frequency

\beq
{6\pi n\over m} = {eB\over m} = \omega_c.
\eeq
The oscillator hamiltonian may be written as
\begin{equation}
H_{osc} =\sum_{q}^{Q} A^{\dag}(q)A(q)\   \omega_c
\end{equation}
where
\begin{equation}
  A(q) = (a(q) + 6 \pi  i P(q))/\sqrt{12\pi}.\label{osdef}
\end{equation}

We are now ready to discuss the upper limit $Q$ that controls the
number of plasma oscillators.  Given that there are only $n$ electrons
in a plane with a total of $2n$ degrees of freedom, the number of {\em
independent} oscillators is bounded by $2n$. We choose the number of
oscillators to be introduced to be {\em equal to the number of
electrons, i.e., $Q=k_F$}. There are many reasons for this choice,
only one of which is intelligible at this point. Recall that we trying
to describe LLL physics for the low energy sector along with high-energy
cyclotronic physics. The oscillators clearly
correspond to the latter. To pay for them, the particle sector has to
give up $n_0 =n$ degrees of freedom, which is precisely what it takes
to project from the full Hilbert space to the LLL. Although
mathematically any value of $Q$ is allowed (being different choices of
generalized gauges) our choice will prove the most suitable.

Our quest for the ground state of the full interacting problem can be
described in two equivalent ways.

Firstly, we can try a variational
approach using a simple product wavefunction in the enlarged Hilbert
space, initially paying no attention to the constraint. The reason  is the
following.
Suppose one were capable of finding the average of the Hamiltonian in
arbitrarily complicated variational wave functions. Then the one with
the lowest energy in the enlarged Hilbert space will be the true
ground state. However, since the Hamiltonian is gauge invariant, we expect the
ground state to be gauge invariant as well, since we know of no case where
gauge invariance is spontaneously broken.  Thus in the variational quest for
the ground state, we can forget gauge invaraince;  the winner will have that
feature.  Of course, we can
can only carry out the variational procedure for product states, so
all the above considerations can be expected to hold approximately. We
will use for this purpose the ground state of $H_0$ which we can write
down by inspection: the composite bosons all condense into the zero
momentum state with wavefunction $\Psi_{CB} (x) =1$ and the
oscillators occupy their ground states:

\beq
\Psi_{osc} =\exp \left[ -\sum_{q}^{Q} { 1\over 12\pi}a^2 ({\bf q})
\right]
\eeq
  giving us  the  MR  wavefunction
\begin{equation}
\Psi^{MR} (x,a) =\exp \left[ -\sum_{q}^{Q} { 1\over 12\pi}a^2 ({\bf
q})
\right] \cdot 1\label{psimr}
\end{equation}

Not only does this minimize $H_0$, both $H_{I}$ and $H_{II}$ have zero
average in this state: Thus this wave function minimizes
$H_0+H_I+H_{II}$ among product wavefunctions. In addition, any density-density
interaction (which
will go as $a^2q^2 V(q)$ since $\rho (q) \simeq q a$ by the
constraint) and $H_{sr}$ will make some difference only at very short
distances, where the wavefunction will be found to have excellent
correlations that keep the particles apart.

Equivalently, we can think of the product wavefunction as a zeroth
order starting point for a perturbative solution based on $H_0$. From
this vantage, the approximation has a chance of being relevant only in
the presence of electron-electron interactions: without them there is
no unique ground state in the exact solution and the product ground
state of $H_0$ must necessarily get destabilized by the neglected
terms. Thus interactions are needed for our approach to work, even
though they are not being explicitly treated.  Note also that $H_0$ does
not correspond to free electrons; for that we need all of $H$ in
Eqn. (\ref{eq-ham}).  By dropping all terms but $H_0$, we have gone
from the free problem with degeneracies to one with a unique ground
state.

There is just one problem: the wave function in Eqn. (\ref{psimr}) is
in the enlarged Hilbert space, whereas the physical wavefunction must
be constrained to lie in the physical subspace. We implement the
constraint by simply multiplying the above wavefunction by a delta
function of the constraint, i.e., by projection.  As explained in the
steps leading up to Eqn.(\ref{psiformulaMR}), $\Psi_{CS}$ is obtained
by setting $a(q) = 6\pi \rho (q)/q$ in $\Psi^{MR} (x,a) $:
\begin{eqnarray}
\Psi_{CS} &=& \exp \left[ -\sum_{q}^{Q} { 3\pi}:\rho    ({\bf q}):{1
\over
q^2}:\rho
(-{\bf
q}):  \right]\label{3pi}\\
&=& \exp \left[ {3 \over 2} \int dx \int dy (\sum_{i }\delta (x-x_i)
-n) \ln
|x-y| (\sum_{j}\delta (x-x_j) -n) \right] \\
&=& \exp \left[ {3 \over 2}  (\sum_{i,j} \ln |x_i -x_j | -2n \sum_{i}
\int dx
\ln  |x-x_i| + \ constants ) \right] \\
& = & \prod_{i<j}|z_i -z_j|^3 \exp \left[ -\sum_j
|z_j|^2/4l_{0}^{2}\right] .
\label{corr}
\end{eqnarray}
  The steps connecting to the penultimate and ultimate lines are from
Kane {\em et al}\cite{kane}.  The integral over $x$ in the penultimate
line may be interpreted as the potential energy of a point charge at
$x_i$, due to a uniform charge density $4\pi n$ and the
two-dimensional Coulomb potential $V(|x -y|) = -{1\over 2\pi} \ln
|x-y|$. (To find the potential at $x$, it helps to first find the
field by invoking the two-dimensional Gauss's law $\oint {\bf E} \cdot
{\bf dr} = \int d^2x \rho (x)$ and then integrate.) Putting back the
phase factors from Eqn.(\ref{eq-phase}), gives us Laughlin's $\Psi_{1/
3}$. \footnote{ Due to the cut-off $Q$, the form of $\Psi$ we get is good only
for $|z_i -z_j| >> 1/Q$.  To continue this form all the way in to get a triple
zero,  we must  invoke the LLL condition.}

 Let us recapitulate. We began by showing that given an exact gauge invariant
eigenfunction
of the enlarged hamiltonian, we could get the physical one by dropping
the delta function of the constraint that it must necessarily have as a
prefactor ( Eqns.  (\ref{psiformula}) and (\ref{psiformulaMR}). ) In the
end we found an approximate variational wavefunction $\Psi^{MR} (x,a)$
that one could argue had a good energy, but was not gauge invariant,
i.e., did not vanish outside the constraint. We then projected it to
the physical subspace by appending the delta function $\delta (\chi )$
($\chi $ being the constraint) and evaluating the product wavefunction
on the constrained  subspace to get $\Psi_{CS}$. Could not multiplication
by $\delta (\chi )$ ruin the good energetics? In general yes, but not
here: given that $\Psi^{MR} (x,a)$ is an approximate eigenstate of the exact
$H$, i.e., obeys the wave equation

\beq H \Psi^{MR}(x,a) \simeq E
\Psi^{MR}(x,a),
\eeq
we can {\em pre}multiply both sides by $\delta (\chi (q))$, {\em
commute it through $H$} to verify that $\delta (\chi (q))
\Psi^{MR}(x,a)$ obeys the same equation.

Our trick for extracting the ground state can be extended to find a
few other states as follows. Consider the sector in which there is a
vortex at the origin. Among all such electronic wavefunctions there
must be one with lowest energy, the ground state in this sector. Let
us then write electronic wavefunction as

\beq \Psi_{e}^{vortex} = \prod_j e^{i\sum_j
\theta_{j}}\Psi'_{e}
\eeq
and subject $\Psi^{'}_{e}$, which is free of
this vortex, to the previous CS transformation. Instead of
Eqn.(\ref{extendedH}) we will end up with

\beq H = {1 \over 2m} \sid
(-i\nabla + e{\bf A}^* + {\bf a}+ (\nabla \times)^{-1} 6\pi \rho +
{{\bf e}_{\theta} \over r} )^2\psi \label{extendedHvort}
\end{equation}
the last term in brackets being due to the vortex prefactor.

The unitary transformation,  which  must now  get rid of the last two
terms in
the bracket, is
\beq
U = \exp [\sum_{q}^{Q} i P(-q)({6\pi \over q} \rho (q) + {2\pi \over
q})].
\eeq

The bosonic wavefunction will again be unity since the external flux
is zero  on  average, including the point flux tube at the
origin. The oscillator wavefunction will still be the gaussian, \beq
\Psi_{osc} =\exp \left[ -\sum_{q}^{Q} { 1\over 12\pi}a^2 ({\bf q})
\right] \eeq but the constraint will now be modified to \beq a(q) =
{6\pi \rho (q) \over q} + {2\pi \over q}.  \eeq because of the
modified $U$.

It is clear that we have now placed a phantom charge of size $1/3$ at
the origin. The plasma will screen it and produce a correlation hole
of charge $-1/3$. This is verified by evaluating the corresponding
wavefunction using the same procedure as before to obtain
 \beq
\Psi_{CS}^{'}(x,
\eta) = \prod_{j}|z_j -\eta | e^{-|\eta |^2/4l_{0}^{*2}} \
\prod_{i<j}|z_i -z_j|^3 \exp \left[ -\sum_j |z_j|^2/4l_{0}^{2}\right]
\eeq
 where we have displayed the answer for the case in which  the vortex
is at  $\eta $ rather than the origin,
to show how even the gaussian factor associated with its location
appears naturally. Note that the magnetic length associated with it
obeys $l_{0}^{*2} = 3l_{0}^{2}$,
appropriate that of a charge $-1/3$
object. Going back to the electronic wavefunction, $|z_j - \eta |$
will become $z_j -\eta $ giving us Laughlin's quasihole state, again with the
understanding that we continue down to $z -\eta =0$ using the LLL condition.

To create two quasiholes, at $\eta_1$ and $\eta_2$,  we do the
obvious
extension of the preceding and find
\beq
\! \! \! \! \! \! \! \!\! \! \! \!\! \! \! \Psi_{CS}^{'}(x,\eta_1, \eta_2) =
|\eta_1 -\eta_2
|^{1/3} \  e^{-(|\eta _1 |^2 + |\eta _2 |^2)/4l_{0}^{*2}}
\prod_{j}|z_j
-\eta_1 |  \ \prod_{j}|z_j -\eta_2 | \ \ \prod_{i<j}|z_i -z_j|^3 \exp
\left[
-\sum_j
|z_j|^2/4l_{0}^{2}\right]
\eeq
Observe that the wavefunction
contains the normalization factor $ |\eta_1 -\eta_2 |^{1/3}$ of the
two-quasihole state. If we follow Halperin\cite{halperin2} and drop
the $mod$
sign (by a singular gauge transformation) we can get a pseudo
wavefunction for
the quasiholes and infer their fractional statistics.

It is possible to assign operators that create the quasiholes in the
following
sense.
Consider the action of the operator
\begin{equation}
\psi (0 ) = \exp \left[  \sum_{\bf q} {2\pi i \over q} P({ -q})
\right].
\end{equation}
on the unprojected ground state which we denote by
\beq
|Osc =0,\Psi_{CB} =1\rangle .
\eeq
 Its action is one of translation in $a$:
\beq
e^{- a^2(q)/12\pi} \to e^{- [ a (q) + {2\pi  \over q} ]^2/12\pi }
\eeq
If we now obtain the projection,  $\wp  \psi (\eta )  |Osc
=0,\Psi_{CB}
=1\rangle
$: we end up with
\beq
\Psi_{CS}^{'} (x, 0) = \prod_{j}|z_j  |  \ \prod_{i<j}|z_i -z_j|^3
\exp \left[
-\sum_j
|z_j|^2/4l_{0}^{2}\right] .
\eeq

It follows that the operator which creates a quasihole at the point
$\eta $  is
\beq
\psi_{qh}(\eta )  = e^{i\sum_j \theta_{j, \eta}}\exp \left[
\sum_{\bf q} {2\pi
i \over q} P({ -q}) e^{iq\eta }
\right].
\eeq
where $\theta_{j, \eta}$ is the angle between the x-axis and the
vector
connecting $\eta$ to $z_j$.

Note that the projection is done only at the very end. In other
words, to
create two quasiholes, we need
\beq
\wp \psi_{qh}(\eta_2) \ \psi_{qh}(\eta_1)\  |Osc =0,\Psi_{CB}
=1\rangle
\eeq
and not
\beq
\wp \psi_{qh}(\eta_2) \wp \ \psi_{qh}(\eta_1)\  |Osc =0,\Psi_{CB}
=1\rangle
\eeq
 In fact  $\wp \psi_{qh} \wp $ has zero matrix elements because
$\psi_{qh}$ is
not gauge invariant.

The adjoint operator $\psi_{qh}^{\dag} (\eta ) $ will append to the
ground
state wavefunction the factor
$\prod_{j}(z_j -\eta )^{-1}$ which will indeed have a charge $1/3$.
However
the resulting wavefunction does  not belong to the LLL  since
negative powers
of $z$ are not allowed. Thus our formalism does  not solve  this
problem with
quasiparticles, first encountered by  Laughlin.

 Let us summarize what has just been done. In the CS representation, the
hamiltonian did
not contain any spurious degrees of freedom, but was very nonlinear
due to the $(\nabla \times )^{-1}$ terms that corresponded to the
nondynamical field ${\bf a}$. We moved to the MR in which the
oscillator degrees of freedom were introduced.  The hamiltonian, which
now contained new canonical degrees of freedom, but no long-distance
($q<Q$) nondynamical fields, had invariance under time-independent
gauge transformations. Physical states had to be gauge singlets to
compensate for the new degrees of freedom.  We found an approximate
variational product ground state $|Osc =0,\Psi_{CB} =1\rangle$, which
was the ground state of $H_0$, a nongauge invariant part of $H$ with a
unique ground state that could be read off by inspection. We then
projected this ground state using $\wp$ to get Laughlin's wavefunction
$\Psi_{1/ 3}(x)$.  This procedure succeeds only because of
particle-particle interactions which render the true ground state
unique and nondegenerate. Continuing, we saw that $\psi_{qh}$ acting on
$|Osc =0,\Psi_{CB} =1\rangle$ produced states, which, upon projection,
led to normalized quasiparticle states. It appears that there is a
dictionary between the eigenstates of $H_0$ and the {\em full
interacting hamiltonian}:

\begin{eqnarray}
|Osc =0,\Psi_{CB} =1\rangle  &\stackrel{\wp}{\rightarrow} &\Psi _{1/
3}(x)\\
\psi_{qh}   (\eta_k )\cdots \psi_{qh}(\eta_1 )|Osc =0,\Psi_{CB}
=1\rangle
&\stackrel{\wp}{\rightarrow} &\Psi _{1/3, \ qh}(x,\eta_1, \cdots \eta_k
)
\end{eqnarray}

All these states may be viewed as ground states in sectors with a
given number of vortices. We do not know if the correspondence carries
over to excited states.

We can now understand  how we have managed to get the phase and
modulus of
$\Psi_{1/ 3}$ in the middle representation. Recall
 Eqn.(\ref{cstocb})
\beq
\psi_{CB}^{\dag} (x) = \psi_{CS}^{\dag} (x) \exp [\sum_{q}^{Q} i
P(-q){6\pi
\over q}  e^{-iqx} ] .\label{cpcs}
\eeq

Thus $\psi_{CB}$ creates a CS boson and introduces a triple quasihole
of charge $-1$, {\em thereby creating the extra particle along with
its correlation hole}. Starting from Halperin's original observation,
all physical pictures, particularly Read's\cite{read1,read2}, have
emphasized that electrons bind to {\em zeros} of the wavefunction, not
just the vortices in the phase, i.e, the flux tubes. {\em In fact only
a zero with an accompanying charge deficit will attract an electron.}
This is the reason we prefer to use the term Composite Boson for the
particle created by $\psi_{CB}^{\dag}$ rather than by
$\psi_{CS}^{\dag}$. The first factor of $\psi_{CB}^{\dag}$carries the
flux, i.e., the phase, and the second, the magnitude, i.e., the
correlation zeros, of $\Psi_{1/ 3}$. (The latter has the effect
of introducing the hole only upon projection.)  Crucial to the success
of all of this is the assignment of independent degrees of freedom to
collective coordinates, rather than trying to express them in terms of
particle coordinates. It is only after $a(q) $ became a canonical
coordinate whose conjugate momentum $P$ could be used to translate it,
that we could so readily produce the correlation hole.

The preceding arguments suggest that $\psi_{CB}$ is a
candidate for being Read's operator\cite{read2}.  Let recall his reasoning for
the
choice of condensate. If we take a Laughlin state with $N+1$ electrons
and kill one at $z$, it looks like an $N$ particle state with a unit
charge correlation hole there. Schematically

\beq \langle N+1 |
\psi_{electron}^{\dag} (z) U(z)^3 |N\rangle \ne 0
\eeq
where $U(z)$ in first quantization appends a factor $\prod_i(z-z_i)$,
i.e., creates a quasihole at $z$. Thus $\psi_{electron}^{\dag} (z)
U(z)^3$ condenses. Given the phase relationship between
$\psi_{elelctron}$ and $\psi_{CS}$, this means $\psi_{CS}^{\dag} (z)
|U(z)|^3$ condenses.  Comparing to Eqn.(\ref{cpcs}) suggests
$\psi_{CB}$ is the Read operator. There is, however, the caveat that
it acts in the enlarged Hilbert space rather than the physical
subspace. While it is true that

\beq
\wp \exp [\sum_{q}^{Q} i P(-q){6\pi \over q} e^{-iqz} ] |Osc
=0,\Psi_{CB} =1\rangle = \prod_j |z-z_j|^3 \Psi_{1/ 3} = |U(z))|^3
\Psi_{1/ 3}
\eeq
it is not true that $\exp [\sum_{q}^{Q} i
P(-q){6\pi \over q} e^{-iqz} ]$ acting directly on the physical state
$\Psi_{1/ 3}$ produces the correlation hole. Thus it is the preimage
of $|U|^3$ (in the big space, before projection), but not an operator
which acting on $\Psi_{1/ 3}$ produces this factor.

There is another related way to say this. As far as $H_0$, which is a
nongauge
hamiltonian goes, it is indeed true that
\beq
\langle Osc =0,\Psi_{CB} =1| \psi_{CB}^{\dag} (\eta_2 ) \psi_{CB}
(\eta_1 )
|Osc =0,\Psi_{CB} =1\rangle \to 1 \ \ as\ \
{|\eta_1 -\eta_2| \to \infty} .
\eeq
On the other hand,  within the gauge theory defined by $H$,
$\psi_{CB}$ is a
gauge dependent operator which cannot have a mean value  in the true
gauge
invariant ground state. (Introducing the line integral between the
points
$\eta_1$ and $\eta_2$ does not help;  it makes it the correlator of
$\psi_{CS}$,
which has the algebraic order found by Girvin and
MacDonald\cite{gm}\cite{fradbook}. )

\subsection{$\nu ={1\over 2}$}
For $\nu ={1 \over 2}$ we choose $l=2$ flux quanta, cancel the
external field, and end up with fermions in zero field on the
average. An analysis of $H_0$ as in the previous subsection will give
us (upon putting back the phase factors of the CS transformation)

\begin{equation}
\psi_{1/ 2} =  \prod_{i<j}(z_i -z_j)^2  \exp \left[ -\sum_j
|z_j|^2/4l_{0}^{2}\right]\cdot |FS\rangle \label{rezread}
\end{equation}
which differs from the Rezayi-Read\cite{rezayi-read} wavefunction by the
absence of projection to the
LLL. This issue will be addressed in the next section.

\subsection{The Jain series $ \nu ={p \over 2p+1}$}

 In this case we we follow Jain\cite{jain-cf} and attach two units of
flux per particle to reduce $e{\bf A}$ down to $e{\bf A}^*$. Since
${\bf A}$ corresponded to $1/\nu$ flux quanta per particle and ${\bf
A}^*$ to $1/p$ quanta per particle

\beq
{A^* \over  A} = {1 \over 2p+1}.
\eeq
Now $H_0$ takes the form
\beq
H_0 = {1 \over 2m} |(-i\nabla + e{\bf A}^*)\psi|^2 +{n \over 2m}(a^2+
(4\pi
P)^2 ).  \label{hjain}
\eeq
First consider  the oscillators. They  are not at the right
(cyclotron)
frequency but at
\beq
\omega_0 = {4\pi n\over m} = {2p \over 2p+1}\omega_c  =
{2\nu}\omega_c .
\eeq
Nonetheless they lead to good  wavefunctions.
Their ground state wavefunction, upon projection,  gives a factor
\beq
\prod_{i<j} |z_i -z_j|^2 \exp \left[- \sum_j {2\nu |z_j|^2\over
4l_{0}^{2}}\right].
\eeq
The derivation  proceeds exactly as in the case of $\nu =1/2$, except
that  we
replace  $4\pi n$ by
$2\nu / l_{0}^{2}$.

The  particles fill $p$ Landau levels. Let us write their
wavefunction as
\beq
\Psi_{CF} = \chi_p = \exp \left[- \sum_j {|z_j|^2\over
4l_{0}^{*2}}\right]
f_p(z,z^*)
\eeq
where $f$ is a polynomial and $l_{0}^{*2} = (2p+1) l_{0}^{2}$ . This
leads to a
wavefunction
\beq
\Psi_{CS} = \prod_{i<j} |z_i -z_j|^2 \exp \left[ -\sum_j {
|z_j|^2\over
4l_{0}^{2}}\right] f_{p} (z,z^*).
\eeq
{\em It is very gratifying that  the two gaussian factors combine
very
naturally to give the right magnetic length} because of the relation
\beq
2\nu + {1 \over 2p+1} = 1.
\eeq

Note that there is no further projection to the LLL. (Jain has pointed out that
even without the projection,  the wavefunction is primarily in the LLL in the
sense that the kinetic energy per particle is only slighly higher than $\hbar
\omega_c$. )
In any event we consider the question of projection to the LLL next.

\section{Transformation to the Final Representation}

The MR has yielded many wavefunctions with good correlations. There are,
however, some weaknesses. First of all, the particles have too much
kinetic energy: in the absence of Coulomb interactions, they should
have no kinetic energy:  $1/m^* =0$. There is as yet no evidence of this
mass renormalization. The oscillators, on the other hand, have  the
right frequency only when $A^*=0$, i.e., for $\nu =1/(2s+1),1/2$.
(Despite all this, the wavefunctions are good because the particle
mass and oscillator frequency do not enter them, only the magnetic
length does. As a result, as long as there is some nonzero $1/m^*$ in
the end, these wavefunctions remain good. In the absence of
interactions, $1/m^*$ indeed vanishes and the nondegenerate ground
state of $H_0$ becomes irrelevant. However, with
interactions, it can be a good approximation to the true ground state).

Another problem has to do with the charge of the particles. The
composite fermions are supposed to have a charge $e^* = e/(2p+1)$ for
$\nu = p/(2p +1)$\cite{laugh,halperin1,halperin2,haldane,jain-cf}. One way to
obtain this result is to imagine injecting a composite fermion by
adding an electron and adiabatically injecting the two quanta of
flux. During the injection, there is tangential electric field. While
the physics may be complicated near the fermion, on a circle of large
radius the effect of this field is to produce a radial current
determined by the bulk Hall conductance. The integrated charge outflow
to infinity or to the edge, i.e., the charge  of the correlation hole, is
readily found to be $-2\nu e$.  Adding to the charge of the injected
electron, the composite fermion charge becomes

\beq e^* = e(1-{2p
\over 2p+1}) = {e \over 2p+1} .
\eeq
The same argument predicts  that the composite boson in $\nu ={1/ 3}$ is
neutral since in that case we add three flux quanta and $3\nu$
electronic units are driven to infinity.

For $\nu =1/2$  these arguments imply  neutral composite fermions.
Read\cite{read1,read2} gives more specific details based on the analysis of
\beq
\psi_{RR} ={\cal{P}}   \prod_{i<j}(z_i -z_j)^2  \exp \left[ -\sum_j
|z_j|^2/4l_{0}^{2}\right]\cdot |FS\rangle .\label{rezread}
\end{equation}
where $|FS\rangle$ denotes the Fermi sea.
Without the $|FS\rangle$ factor, each electron sits in the middle of the
double correlation hole and the complex is neutral. The role of the
Slater determinant is to produce the antisymmetry. It contains factors
of the form  $e^{i{\bf k} \cdot {\bf r}} = e^{i( kz^* + k^*z)/2}$. Since $z^*$
acts like $2 l_{0}^{2} {\partial \over \partial z} $ upon projection
by $\cal{P}$\cite{jach}, its action on the Jastrow factor is to split the
double
zero and move the electron off the center of the correlation hole. The
shift in $z$ is $ikl_{0}^{2}$ and the dipole moment associated with a
fermion of momentum $k$ is $e l_{0}^{2}\ \hat{{\bf z}} \times {\bf
k}$. Note that projection to the LLL is essential to the above
picture.

The reason all these features are not evident in the MR is that the
particles are coupled to the oscillators and neither is a true
quasiparticle with its ultimate (or nearly ultimate) mass, charge
etc. One way to deal with this is to integrate out one in favor of the
other. {\em We take the more symmetric route of eliminating the
coupling between them.}This will be done within the following
approximation scheme:
\begin{itemize}
\item We will work at long distances. Thus if any quantity has an
expansion in powers of $q$ we will keep just the leading term. This
involves the neglect of derivative couplings.
\item When we encounter the density operator in a product with other operators,
we will use the RPA:
\beq
\sum_j e^{i(q-k))x_j} \simeq n (2\pi )^2 \delta^2 ({\bf q-k}).
\eeq

\end{itemize}
 Thus we will  ignore
$H_{sr}$, the short range piece due to $\delta {\bf a}$, and the non-RPA term
$H_{II}$.

The analysis of $\nu = {p \over 2p+1}$ will be carried out in first
quantization. The particles are fermions.

The hamiltonian $H_0 + H_I$
reads
\beq
\!\!\!\!\!\!\!\!\!\!\!\!H  =\sum_j{\Pi_{j}^{2}\over 2m}  +
\sum_{q}^{Q}
A^{\dag}(q) A(q) \omega_0
+ \theta \omega_0 \sum_{q}^{Q} \left[ c^{\dag }(q) A(q) + A^{\dag}
(q) c(q)
\right] \equiv  T + H^{osc} + H_I\label{fqham}
\eeq
where
\begin{eqnarray}
{\bf \Pi } &=& {\bf p} + e{\bf A}^*\\
\nabla \times e{\bf A^*} &=& -eB^* = -{eB\over 2p+1}\\
A(q) &=& { 1 \over \sqrt{8 \pi}} [ a(q) + 4\pi i P(q) ]\\
\theta &=& {\sqrt{2\pi }\over 4\pi n}\\
\omega_0 &=& {4\pi n\over m} = {2p \over 2p+1}\ \omega_c \\
c(q) &=& \hat{q}_{-} \sum_j \Pi_{j+} e^{-iqx_j}\\
V_{\pm} &=& V_x \pm iV_y\\
\left[ A(q),A^{\dag} (q') \right] &=&  (2\pi )^2 \delta^2({\bf q}-{\bf q}')
\end{eqnarray}

  It is useful to know that
\beq
\left[ \Pi_- , \Pi_+ \right] = -2eB^*, \label{picom}
\eeq
and that in our approximation
\begin{eqnarray}
\left[  c(q),c^{\dag} (q')\right] &=& \hat{q}_-' \hat{q}_+\sum_j \left[
\Pi_{+}^{j} , \Pi_{-}^{j} \right]
e^{-i(q-q')x_j} + {\cal{O}}(q) \\
&\simeq & 2eB^{*}n (2\pi )^2 \delta^2({\bf q}-{\bf q}').
\end{eqnarray}

The other commutators vanish to this order in $q$:
\beq
\left[ c(q),c (q')\right] = \left[ c^{\dag}(q),c^{\dag} (q')\right] =
{\cal{O}}(q,q').
\eeq

The coupling $H_I$ will be eliminated by a canonical transformation

\beq U(\lambda_0 ) = e^{iS_0 \lambda_0 } =\exp \left[ \lambda_0 \theta
\sum_{q}^{Q}(c^{\dag} (q) A(q) - A^{\dag}(q)c(q)) \right]
\eeq
where
$\lambda_0$ is to be chosen appropriately.  The operators of the MR, called
$\Omega^{old}$, will be related to those of the FR (with no superscripts) by

\beq
\Omega^{old} = e^{-iS_0\lambda_0 } \Omega e^{iS_0\lambda_0 }.
\eeq

We will also define

\beq \Omega (\lambda ) = e^{-iS_0\lambda } \Omega
e^{iS_0\lambda } \eeq so that \beq \Omega (\lambda_0) = \Omega^{old} \
\ \ \ \ \ \ \ \ \ \ \ \ \ \ \Omega (0) =  \Omega
\eeq

We start with the flow equations
\begin{eqnarray}
{dA(q,\lambda )\over d\lambda} &=& -\theta c(q,\lambda )\\
{dc(q,\lambda )\over d\lambda} &=& 2eB^*n \theta A(q, \lambda ).
\end{eqnarray}
{\em It is understood that $0<q\le Q$ in the above and what follows.}
The flow equations have  the solution
\begin{eqnarray}
A(q,\lambda ) &=&  \cos \mu \lambda \ A(q) - {\theta \over \mu} \sin \mu
\lambda \
c(q)\\
c(q,\lambda ) &=&  \cos \mu \lambda \ c(q) + {  \mu \over \theta} \sin
\mu
\lambda \ A(q)\\
\mu^2 &= & 2eB^*n\theta^2 = {1 \over {2p}}.
\end{eqnarray}
Consider the kinetic energy, which may be written, using
Eqn.(\ref{picom})  as
\beq
T= \sum_j {\Pi_{-}^{j}\Pi_{+}^{j}\over 2m} + \sum_j {eB^*\over 2m}.
\eeq
The second term, which  does not evolve to the order we are working in, will
turn out to describe the magnetic moment coupling of SSH\cite{SSH}.
The first  evolves as per
\beq
{dT \over d\lambda} = {eB^*\theta \over m} \sum_{q}^{Q} (A^{\dag}
(q,\lambda )
c(q,\lambda ) + c^{\dag}(q,\lambda ) A(q,\lambda ))
\eeq
so that
\beq
\!\!\!\!\!\!\!\!\!\!\!\!\!\!\!T(\lambda ) =T+ {eB^*\theta \over m}
\sum_{q}^{Q}
\left[ {\sin 2\lambda \mu \over 2\mu } (c^{\dag}(q) A(q) +
A^{\dag}(q) c(q)) +
{1 - \cos 2 \mu \lambda \over 2 \mu } ({\mu \over \theta}
A^{\dag}(q)A(q) -
{\theta \over \mu}c^{\dag}(q)c(q))\right]
\eeq
The oscillator hamiltonian and $H_I$ assume the following form in terms of FR
operators:
\beq
\!\!\!\!\!\!\!\!\!\!\!\!\!\!\! H^{osc} = \omega_0 \sum_{q}^{Q}
\left(A^{\dag}(q)A(q) \cos^2 \mu \lambda -(A^{\dag}(q)c(q)
+c^{\dag}(q)A(q))
{\theta \over 2\mu} \sin 2\mu \lambda + {\theta^2 \over \mu^2} \sin^2
\mu
\lambda c^{\dag}(q)c(q)\right)
\eeq
and
\beq
\!\!\!\!\!\!\!\!\!\! H_I = \theta\omega_0 \sum_{q}^{Q} \left[ \cos2
\mu \lambda
  (c^{\dag}(q)A(q) + A^{\dag}(q)c(q)) + \sin 2\mu \lambda ({\mu \over
\theta}
A^{\dag}(q)A(q) - {\theta \over \mu}c^{\dag}(q)c(q))\right] .
\eeq
We will now choose $\lambda =\lambda_0$ such that cross terms between
the
particles and oscillators vanish. Gathering the coefficients of
$A^{\dag} (q)\
c(q)$ from $T,H^{osc}$ and $H_I$  we find that they add to zero if
\beq
\tan \lambda_0 \mu = \mu = {1 \over \sqrt{2p}} \label{bestlambda}
\eeq
This completely fixes the canonical transformation.

We now ask what the frequency of the oscillators is and find that the
coefficient of the $A^{\dag}A$ term is exactly $\omega_c = eB/m$!
This agreement  with Kohn's theorem  confirms the soundness of our decoupling
transformation whose only free parameter has already been chosen.

What about the particles? The $c^{\dag}c$ term has coefficient $- {1
\over
2mn}$. Thus the total particle kinetic energy is
\beq
T = \sum_j {\Pi_{-}^{j}\Pi_{+}^{j}\over 2m} +\sum_j {eB^*\over 2m} -
{1 \over
2mn}\sum_i\sum_j \sum_{q}^{Q}
\Pi_{-}^{i}e^{-iq (x_i-x_j)} \Pi_{+}^{j}. \label{kinetic}
\eeq
We also need to transform the constraint
\beq
\rho^{old} (q) = {qa^{old}(q)\over 4\pi}
\eeq
on physical states by transforming both sides of the equation.  The  flow
equation
\beq
{d \rho (q,\lambda )\over d\lambda } = {q \over
\sqrt{8\pi}}(A(q,\lambda )+
A^{\dag} (-q,\lambda ))
\eeq
can be integrated to give
\beq
\rho^{old}(q) = \rho (q) + {q \over \sqrt{8\pi}}\left( {\sin \mu
\lambda_0\over
\mu} (A(q)+ A^{\dag} (-q)) - {\theta \over \mu^2}(1- \cos \mu
\lambda_0
)(c(q)+c^{\dag}(-q)) \right)\label{newrho}
\eeq
while  previous results for $A$ tell us that
\beq
{qa^{old}(q)\over 4\pi} = {q \over \sqrt{8\pi}}\left[ {\cos \mu
\lambda_0}
(A(q)+ A^{\dag} (-q)) - {\theta \over \mu}\sin \mu \lambda_0
(c(q)+c^{\dag}(-q))\right] .
\eeq
{\em Remarkably, the terms involving the oscillators match on both
sides upon
using $\tan \mu \lambda_0 = \mu$ and the constraint involves only the
particles}:
\beq
\!\!\!\!\!\!\!\!\!\!\!\!\rho (q) =  - {i \l_{0}^{2}( 1 - \cos \mu
\lambda_0)
\over \cos \mu \lambda_0
\sin^2 \mu \lambda_0}\sum_j (q \times \Pi_j )e^{-iqx_j}
= - {i \l_{0}^{2} \over \cos \mu \lambda_0 ( 1 + \cos \mu \lambda_0)}
\sum_j (q \times \Pi_j )e^{-iqx_j}
\eeq

This is very fortunate since it is no use decoupling the particles in
the hamiltonian if the constraint still couples them.

Now we  calculate   the transformation of the   current

\beq
J_{+}^{old} (q) =  \sum_j  \left[ {\Pi_ j\over m} e^{-iqx_j} + {n
\over m}
\sqrt{8\pi} \hat{q}_+ A(q)\right]^{old}
\eeq
to  the  FR. (Note that the non-RPA part of  it has been dropped). We
find

\beq
J_+(q) = {\hat{q}_+ \omega_0 \over \sqrt{2\pi} \cos \mu \lambda_0 }
A(q) = {\hat{q}_+ \omega_c \ \cos \mu \lambda_0 \over \sqrt{2\pi}  }
A(q).
\eeq

Remarkably the entire contribution comes from the oscillators; the
part proportional to $\Pi$ just cancels out.

To summarize, in the  FR,  for the fraction $\nu = \ p/(2p+1)$
\begin{eqnarray}
\!\!\!\!\!\!\! H &=&\!\!  \sum_j {\Pi_{-}^{j}\Pi_{+}^{j}\over
2m} \!
+\sum_j {eB^*\over 2m} - \!\! {1 \over 2mn}\! \sum_i\sum_j
\sum_{q}^{Q}
\Pi_{-}^{i}e^{-iq (x_i-x_j)} \Pi_{+}^{j} \!+ \!\sum_{q}^{Q} \!
A^{\dag}(q)A(q)
\omega_c \label{nonintham}\ \ \ \\
\chi (q) &=& \rho (q)+ {i \l_{0}^{2} \over \cos \mu \lambda_0 ( 1 +
\cos \mu
\lambda_0)}
\sum_j (q \times \Pi_j)e^{-iqx_j} =0 \ \ \  (constraint)\label{cons}
\\
J_{+}^{old}(q)  &= &{\hat{q}_+ \omega_0 \over \sqrt{2\pi} \cos \mu
\lambda_0 }
A(q) =(J_{-}^{old}(-q))^{\dag}\\
\rho^{old}(q) &=&  {q {\cos \mu \lambda_0} \over \sqrt{8\pi}} (A(q)+
A^{\dag}
(-q)) +\rho (q)  - {il_{0}^{2} \over 1 + \cos \mu \lambda_0}  (\sum_j
(q \times
\Pi_j)e^{-iqx_j} ) \label{rold}\\
\tan \mu \lambda_0 &=& {1 \over \sqrt{2p}}=\mu
\end{eqnarray}
where once again we are assuming $0<q\le Q$. Outside this region
there is no
difference between old and new variables. Note that the
transformations do not
require that $Q$ have  any particular value\cite{bert}.

Consider Eqn.(\ref{nonintham}) for the hamiltonian, focusing on the
particle sector.  We see that decoupling the oscillator has lead to
the third term. In it is a one-particle piece corresponding to $i=j$,
those with $i\ne j$ being additional interactions. When we combine this $i=j$
term
 with the first term, we see the mass gets renormalized to :

\beq
{1 \over m^*} = {1 \over m} ( 1- {1\over n}\sum_{q}^{Q})
\label{massren}.
\eeq

As the upper limit $Q$ grows, the particle mass increases and the
kinetic energy gets quenched. In the noninteracting case we are
studying, the kinetic energy must be fully extinguished. {\em Note
that our choice $Q=k_F$, i.e., $n_{osc} =n$ accomplishes this
exactly.} In addition, this is the right number of constraints, all in
the particle sector, to restrict it to the LLL. A smaller choice of
$Q$ gives the particles kinetic energy of order $1/m$ and also does
not impose enough constraints to project to the LLL. One is free to
use a smaller value for $Q$, but one will have a harder time going
from such a hamiltonian to the correct behavior for the noninteracting
case. A larger value $Q>k_F$ leads to a negative $1/m^*$,  clearly a bad
starting point when we go on to introduce interactions. Note
that the kinetic energy of the fermions is quenched independent of
their momentum.  We do not know what set of diagrams, if any,
corresponds to carrying out this canonical transformation. Recently
 Chari, Haldane and Yang\cite{chari} worked on a gauge invariant
approximation scheme for mass renormalization in a theory with flux
tubes of size $1/\Lambda$.  Unfortunately, to leading order in
$\Lambda /k_F$, the mass shift went the wrong way.  Presumably higher
order terms (which are surely needed since the point flux limit is
$\Lambda /k_F \to \infty$) will fix this.

Once the kinetic energy is quenched all that is left of the $i=j$
term is

\beq
\sum_j {eB^*(r_j) \over 2m}.
\eeq
where we have emphasized that $B^{*}$ (a constant so far) is to be
evaluated at $r_j$ as per our calculation.  Suppose we change the
external field by $\delta {\bf A}$. Repeating the analysis with this
field will lead back to the above formula, but with an extra piece
${e\delta B\over 2m}$ per particle. This result also follows if
$\delta B$ varies very very slowly in space, say between one galaxy
and the next. {\em This is exactly the coupling of a magnetic moment
$e/2m$, mandated by SSH\cite{SSH}.} We did not have to attach it by hand, it
emerged naturally. Note that the $m$ in this term did not get
renormalized.  We could also write this term as follows

\beq \sum_j
{e\delta B(r_j) \over 2m} = \int d^2x \  {e\over 2m}\ \ \rho (x) \nabla \times
\delta
{\bf A} = -\int d^2x \delta {\bf A} \cdot {\bf j}_{mag}
\eeq
where
\beq
{\bf j}_{mag} = {e \over 2m} \hat{\bf z} \times \nabla n
\eeq
is the current associated with uncancelled cyclotronic currents\cite{GMP,SSH}.
It follows that if we add such a $\delta A$, it will lead to a charge density
$\delta \rho = K_{00} {e \delta B \over 2m}$.  Taking the ratio of the applied
vector potential  to the induced  $\delta \rho$, we obtain
\beq
K_{01} = iq {e \over 2m} K_{00}.
\eeq
Note that we did not calculate $K_{00}$, but merely  showed that  $K_{01}$ has
a piece proprtional to it  with a factor $ {iqe \over 2m}$.

Let us now turn to  the $i\neq j$ pieces in Eqn.(\ref{nonintham}).We
treat the
$q$ sum as follows:
\beq
\sum_{q}^{Q}\exp{-iq(r_i-r_j)}=\delta^2(r_i-r_j)-\sum_{q>Q}\exp{-iq(r_
i-r_j)}
\eeq

The delta function vanishes on spinless fermion wavefunctions or on
hard-core boson wavefunctions, and can safely be dropped. (Note that
when $H$ acts on a wavefunction to its right, we can move this delta
function through the $\Pi_{+}^{j}$ to get the desired zero since the
exchange produces corrections of higher order in $q$.) We have thus
reduced the $i\neq j$ pieces to a short-range interaction to be lumped
with (the RPA part of) $H_{\delta a}$, the contribution from the short
range CS field $\delta{\bf a}$ that was not made dynamical.  What is
their collective role? We know that they cannot feed back to our
zeroth order hamiltonian since a vanishing kinetic energy for LLL fermions is
an exact result in the noninteracting theory.  On the other hand, in
the case of $\nu =1/2$, we can show that these terms affect the large
$q$ sector in the following very desirable way.  In RPA, the large $q$
fields ({\em including the term we shifted from small to large $q$} )
conspire to produce the plasmon pole with the right position and
residue in the density-density correlation, as shown in Appendix
A. The issue is the following: There are no oscillators for $q>Q$. The
fermion mass has been renormalized to $m^*$, but they still have to
produce the cyclotron pole at $eB/m$ at large $q$ via a collective
mode. One therefore needs a Landau parameter $f_1$ to restore the
correct cyclotron pole\cite{hlr,simon}. The $i\ne j$ interactions (when
shifted to large $q$) fulfil exactly that role.  It is very satisfying
that although our oscillators no longer exist for $q>Q$, the physics
(plasmon pole and residue to order $q^2$ ) is continuous across this
border.

\subsection{Electronic charge density}
Consider next Eqn.(\ref{rold}) for the charge density
\beq
\rho^{old}(q) =  {q \ {\cos \mu \lambda_0} \over \sqrt{8\pi}} (A(q)+
A^{\dag}
(-q)) +\rho (q)  - {il_{0}^{2} \over 1 + \cos \mu \lambda_0}  (\sum_j
(q \times
\Pi_j)e^{-iqx_j} )\label{rhonat}
\eeq
and the operator that is equivalent in the physical subspace
\beq
{qa^{old}(q)\over 4\pi} = {q \ {\cos \mu
\lambda_0}
 \over \sqrt{8\pi}}\left[ (A(q)+ A^{\dag} (-q)) \right] - {il_{0}^{2} \over
\cos \mu \lambda_0}
(\sum_j
(q \times \Pi_j)e^{-iqx_j}.
\eeq
In the MR, any combination of the form
\beq
 \gamma \rho^{old}(q) + (1-\gamma ) {qa^{old}(q)\over 4\pi}\label{x}
\eeq
is an acceptable definition of the physical charge density. (For
example, HLR use this freedom to write the Coulomb interaction
entirely in terms of $a$.) We could canonically transform any such
combination to represent charge density in the restricted physical
space.  In an exact calculation there will be nothing to choose
between them. In our approximation scheme, however, the following
combination stands out:
\begin{eqnarray}
\rho^{preferred} &=& {1 \over 2p+1}  \rho^{old}(q) + {2p \over 2p+1}
{qa^{old}(q)\over 4\pi}\\
&=& {q \over \sqrt{8\pi}} {\cos \mu \lambda_0} (A(q)+ A^{\dag} (-q))+
{\rho (q) \over 2p+1} -{il_{0}^{2} }  (\sum_j (q \times
\Pi_j)e^{-iqx_j}
).\label{rhobar}
\end{eqnarray}
To see what is special about it,
consider $\bar{\rho}$, its  projection  to the LLL, which clearly
corresponds
to the ground state of the oscillators,  and is  obtained by dropping
the terms
linear in $A$ and $A^{\dag}$:
\beq
\bar{\rho}(q) = {\rho (q) \over 2p+1} -{il_{0}^{2} }  (\sum_j
(q \times
\Pi_j)e^{-iqx_j} ).
\eeq

 Now we know from the work of Girvin and Jach\cite{jach} and
GMP\cite{GMP} that the projected density must obey  the algebra of magnetic
translations:
\beq \left[ \bar{\rho}(q) , \bar{\rho}(q') \right] = i
l_{0}^{2} (q\times q') \bar{\rho} (q+q').\label{GMP}
 \eeq
 in the small
$q$ truncation  of the structure constant, a truncation
which  satisfies the Jacobi
identity)\cite{nick}.
 {\em Our $\bar{\rho}$
obeys this algebra.}
There are two points that should be noted in this connection.

\begin{itemize}
\item The magnetic algebra is  expected upon projection to the LLL only
in the original electronic problem, i.e., between physical states.  In
evaluating a product or commutator,  we must therefore restrict intermediate
states to be physical.  The exceptions are operators that do not mix physical
and unphysical states, i.e., gauge invariant ones. These can be multiplied
freely in the full space.  Since we computed the $\bar{\rho}$  commutator with
no restriction on intermediate states, we expect that $\bar{{\rho}}$
is gauge invariant\cite{ginote} and that we can freely multiply our
$\bar{\rho}$ of Eqn(\ref{rhobar}) in the enlarged space without
worrying about constraints.
\item The algebra contains terms up to cubic order in $q$ in the right
hand side. These can arise  from the terms we have kept and
terms we have neglected. {\em The truncation we have chosen has the virtue
that the algebra is satisfied with the terms we have, with no need to
appeal to higher order terms}. Presumably the leading higher order
terms cancel out in this preferred combination.
\end{itemize}

 For these reasons,  we shall  use this $\bar{\rho}(q)$ and the
corresponding
$\rho^{old}$:
\begin{eqnarray}
\rho^{old} &=& {q {\cos \mu \lambda_0} \over \sqrt{8\pi}} (A(q)+
A^{\dag} (-q))
+
{1\over 2p+1}{ \sum_j  e^{-iqx_j} } -il_{0}^{2} \sum_j (q \times \Pi_j )
e^{-iqx_j} \label{rhoold}
\end{eqnarray}
when we need a formula for the electronic charge density. Since the
particle
kinetic energy has vanished, the entire hamiltonian will be given by
the
Coulomb interaction written in terms of the above density. It will
evidently be
gauge invariant.

 {\em Our  formula for the charge tells us  that the CF has charge
$1/(2p+1)$.}
 When $p=\infty$ i.e., $\nu =1/2$ , this part of the charge vanishes
and we
are left with just the second dipolar  piece. The latter  couples to
an
external scalar potential $\Phi $ as per
\beq
H_{ext} = (-e) \sum_q \rho (-q) \Phi (q) = -iel_{0}^{2} \sum_q \sum_j (q
\times p_j
) e^{iqx_j} \Phi (q) = \sum_j -el_{0}^{2} \hat{\bf z} \times p_j \cdot
{\bf
E}(x_j)
\eeq
where ${\bf E}$ is the electric field.
Note the anticipated value of the dipole moment\cite{rezayi-read,read2,dipole}.

We have argued that our charge density is gauge invariant in that it does not
mix physical and unphysical states.\footnote{ Usually we say that an operator
is gauge invariant if it commutes with the constraints, which are the
generators of the gauge transformations. However if we merely require that the
operator not turn physical to unphysical states, it is  sufficient if its
commutator  with the constraints is  proportional to the constraints. The
Faddeev-Popov\cite{fp} method applies to this general case.}   This means
it must
either commute with the  constraints or have a commutator
proportional  to
them.Indeed we find
\beq
\left[ \bar{\rho}(q) ,  \chi(q') \right] = il_{0}^{2} (q \times q')
\chi
(q+q').\label{gi}
\eeq

Our joy is tempered by the algebra of constraints. We find
\beq
\left[ {\chi}(q) ,  {\chi(q')} \right] = il_{0}^{2} {(q' \times q)  \over
\cos^2 \mu \lambda_0}
 \left[      \rho (q+q')+ {i \l_{0}^{2} \over    ( 1 +
\cos \mu
\lambda_0)^2}
\sum_j ((q +q') \times \Pi_j)e^{-i(q+q')x_j}
                                \right] .
\eeq
Note that  the operator in the right hand side is not quite the constraint; the
term of order $(q\times  q')(q+q') $ has the wrong coefficient. In contrast to
the charge algebra which closed and was correct even to  order $(q\times
q')(q+q') $, the constraint algebra closes only to leading order if we use the
truncated expressions. Thus  we have a situation where the Faddeev-Popov method
applies  ( the hamiltonian commutes with the constraints to give  constraints
which in turn form an algebra) only to leading order.

What if  we   use
the projected version of   Eqn. (\ref{rhonat}) for the charge? The reader may
verify that GMP algebra is  satisfied   and  the constraints  commute with the
hamiltonian, both  to leading order ($q\times q'$). In this version the charge
of the CF will not be   transparent, but the conservation laws and Ward
identities will be.
These comments apply to a models  obtained by truncating  operators to the
orders calculated. Non-RPA and higher order terms in $q$ can modify the
commutators\cite{nonrpa}. We do not pursue this issue for  the gapped states
for
which we will propose an approximation scheme that ignores the
constraints.

Of all the results that flowed from the canonical
transformation,
only $1/m^*=0$ requires $Q=k_F$. For example the CF charge
($1/(2p+1)$ ) and dipole moment are
true for any $Q$\cite{bert}.

\subsection{Hall Conductance}
A common question one has for $\nu =1/2$ is who carries the Hall
current if the CF is neutral. The answer, for general $\nu =
p/(2p+1)$, and not just for $\nu=1/2$, is that the oscillators carry
the entire Hall current in the clean limit considered here.  We show this as
follows.
\begin{itemize}
\item We couple the system to an external potential $\Phi(q) $ using
the
formula Eqn.(\ref{rhoold}) for $\rho^{old}$.
\item We recall that in the passage to the  FR, the transport current
due to
particle motion got exactly  canceled so that finally :
\beq
J_{+}^{old}(q)  = {\hat{q}_+ \omega_c \ \cos \mu \lambda_0 \over \sqrt{2\pi}  }
A(q).
\eeq
\item In view of the above, we focus on just the oscillator sector,
find
$\langle A \rangle $  due to the potential and from it, $\langle
J\rangle$.
\end{itemize}

 So we begin with
\begin{eqnarray}
H_{osc} &=& \sum_q \omega_c A^{\dag}(q)A(q) - e \sum_q \Phi(q)
\rho^{old}(-q)
\\
&=& \sum_q \omega_c A^{\dag}(q)A(q) - e \sum_q \Phi(q) {q \over
\sqrt{8\pi}}
(A(-q) + A^{\dag}(q)) \cos \mu \lambda_0\\
&=& \sum_{q} \omega_c \left[ A^{\dag}(q) - {qe \cos \mu
\lambda_0\over
\sqrt{8\pi}\omega_c} \Phi (-q)  \right] \!  \cdot \! \left[ A(q) - {q
e \cos \mu
\lambda_0\over \sqrt{8\pi}\omega_c } \Phi(q)  \right]\!\! +\!\!
const.  \\
\end{eqnarray}
The new ground state of the oscillators is found by shifting $A$ as
follows:
\beq
\langle A(q) \rangle =  {qe \over \sqrt{8\pi}\omega_c }\cos \mu
\lambda_0
\Phi(q)
\eeq
 The ground state electromagnetic current is
\beq
\langle (-e) J_{+}^{old}(q) \rangle = (-e) {\hat{q}_+ \omega_c \ \cos
\mu \lambda_0 \over
\sqrt{2\pi}  } <A(q)> = - {e^2 \over h} q_+ \nu \Phi(q)
\eeq
 where we have reinstated $\hbar = 1= h/2\pi $ to show that we have
the correct value $\sigma_{xy} = \nu e^2/h$.  By translation
invariance (in the clean system) the Hall conductance has to have this
value. What is significant is that it all comes from the
oscillators. In general we can write in our approximation scheme the
following expression for longitudinal and Hall conductivities

\beq
\sigma_{\mu \nu} = \sigma_{\mu  \nu}^{osc} +\sigma_{\mu
\nu}^{particles}
\eeq
since the charge $\rho^{old}$ and hamiltonian separates into pieces,
one for the oscillators and one for the particles, as shown in our letter
\cite{prl}. (The particle part
happened to be negligible for the Hall conductance in the clean
system. When we turn on interactions, the particle current  will acquire a
piece proprotional to $e^2$. )   This additivity of $\sigma_{\mu\ \nu}$,  the
has also been independently   obtained
by D.H.  Lee\cite{dh}, who statrted with the bosonic theory. {\em In contrast,
in the usual CF
fermion theories in which the CF is fully charged, one adds
resistivities.} The concerns of Lee {\em et al}\cite{proto} on the
value of $\sigma_{xy}^{CF}$ therefore do not therefore apply to our
approach.

\section{Introducing interactions}

So far we have been considering just the noninteracting theory. A
unique ground state appropriate to an interacting theory emerged in
the MR upon keeping just $H_0$ and dropping $H_I$ and $H_{II}$. This
led to good wavefunctions, but bad dispersion relations for the
oscillators (wrong frequency at $q=0$ except when $A^*=0$) and a
particle kinetic energy of order $1/m$ instead of zero.  We then dealt
with $H_I$ and $H_{II}$: eliminating $H_I$ by the canonical
transformation and dropping $H_{II}$ as non-RPA. This led to
oscillators of the right frequency and particles with no kinetic
energy and just a magnetic moment $e/2m$ which coupled to variations in the
background $B$ field. These are the correct results for the
noninteracting case.

We are now going to include  interactions,
illustrating the procedure with a Coulomb interaction cut off in the
ultraviolet at $q=Q$ so that it may be described entirely by our
oscillators. (This rounding off of the Coulomb potential at short
distances may be a realistic description of the experimental system
with finite thickness.) The full hamiltonian is:
\begin{eqnarray}
H_{T} &=&  \sum_q \omega_c A^{\dag}(q)A(q) + \sum_i {e B^*\over 2m} +
{1 \over
2} \sum_{q}^{Q}  \rho^{old}(-q) {2\pi e^2\over q}\rho^{old}
(q)\label{htot}\\
\rho^{old}(q) &=&  {q {\cos \mu \lambda_0} \over \sqrt{8\pi}} (A(q)+
A^{\dag}
(-q)) +{\sum_j e^{-iqx_j}\over 2p+1}   - {il_{0}^{2}}  (\sum_j (q
\times
\Pi_j)e^{-iqx_j} ).
\end{eqnarray}

When we expand out $\rho^{old}(-q) \ v(q)\ \rho^{old} (q)$ we will get a piece
that renormalizes the oscillator frequency to order $e^2q$ (which for
short range interactions would have been order $e^2 q^2v(q)$) as per
Kallin and Halperin\cite{Kallin}), a mixing of order $e^2q$ between
oscillators and particles and a $\bar{\rho}(-q) {\pi e^2\over
q}\bar{\rho} (q)$ term in the particle sector. The mixing can be
eliminated by a further canonical transformation, which can be done to
lowest order in $e^2q$.  It will modify all previous formulas for
$\rho^{old}$, $J$ and so on to terms of higher order.  We
ignore these changes in our analysis,  which is limited to lowest
nontrivial order in $q$ and $e^2$.

Before going to the particle sector for a detailed analysis let us
note that if
in the  density-density correlation $K_{\rho \rho}$ we consider just
the
oscillator part of $\rho^{old}$:
\beq
\rho^{old} \simeq {q {\cos \mu \lambda_0} \over \sqrt{8\pi}} (A(q)+
A^{\dag}
(-q))
\eeq
we find the cyclotron pole
\beq
K_{00} (q\ \omega ) \simeq  {q^2 \omega_c\cos^2 \mu \lambda_0  \over
4\pi
(\omega_{c}^{2}-  (\omega +
i \eta
)^2 )}
\eeq
whose residue obeys the sum rule
\beq
\int_{0}^{\infty} Im K_{00} (\omega ) \omega d \omega = {q^2 n \pi
\over
2m}.
\eeq
Since we work in unit volume $n =N$, the particle number.

Turning to the particle sector, the hamiltonian is just

\beq H_{part}
= \sum_{q}^{Q} \bar{\rho}(-q) {\pi e^2\over q}\bar{\rho }(q).
\eeq
(We have dropped the $eB^*/2m$ term assuming that the external field
is homogeneous).

Recall that GMP\cite{GMP} began with the above hamiltonian, the LLL assumption,
and the
magnetic algebra of the $\bar{\rho}$'s, and proceeded to derive the
spectrum of collective excitations in the Single Mode
Approximation. What is new here, as far as low energy physics is
concerned?  The first point is that we have not just the algebra, but
a {\em concrete realization of the algebra in terms of canonical
variables in terms of which we can try to do explicit calculations.}
This is similar in spirit to the recent work of Haldane and
Pasquier\cite{pasquier} who started with an algebraic formulation of
the LLL problem and then switched to a concrete realization of the
algebra in terms of canonical operators. Next,  our approach derives the
representation from first principles, and
shows  how a field theory in the full Hilbert space leads
to LLL physics upon identifying the plasma oscillators and then
freezing them. Finally there are many low energy responses depending on $m$,
which we obtain by starting in the full space and systematically separating
high and low energy degrees of freedom.

The particle  hamiltonian, written out fully,   (dropping the
$e\delta B/2m$
term) is
\begin{eqnarray}
H_{part}&=&\sum_{q}^{Q} \left( {\rho (-q)\over 2p+1}  +il_{0}^{2}
(\sum_j (q
\times \Pi_j)e^{iqx_j} )\right) {\pi e^2\over q} \left( q \to
-q\right)\\
&=& { l_{0}^{4}}\sum_{q}^{Q}  \sum_{i,j} {\pi e^2\over q}(q \times
\Pi_j)e^{iq(x_j-x_i)}(q \times \Pi_i ) \nonumber \\
&+& \sum_{q}^{Q}\sum_{i,j} {\pi e^2\over q} {e^{iq(x_i -x_j)}\over (
2p+1)^2} +
{i l_{0}^{2}\over 2p+1} \sum_{q}^{Q}  \sum_{i,j}\bigl\{  {\pi e^2\over q}
(q \times
\Pi_j)e^{iq(x_j-x_i)} - h.c.\bigr\} \label{partham}
\end{eqnarray}

This hamiltonian affords a concrete, microscopic, and detailed
realization of composite fermions.  Not only does it describe particles
that see the weaker field $B^*$ that is just right to fill $p$ Landau
levels, it also describes particles with local charge $e^* =
e/(2p+1)$. The main attraction of Jain's CF picture is that one can think of
the original problem in terms of roughly noninteracting particles. Since our
description shows the right charge and a mass of Coulombic origin at tree
level, we may expect that only small corrections separate what we see in the
hamiltonian from what finally happens.

 The hamiltonian is,  however,  not that of  free fermions
in the weaker field. Indeed,  this is not to  be expected,  because the
Hilbert space of free fermions is too big to describe the LLL physics
that is left after we have frozen the oscillators.  That is why we
have constraints to reduce the size of the Hilbert space and  the hamiltonian
has to be gauge invariant under the transformations they generate (to the
approximation we are
working in). The free particle hamiltonian does not have this property.

However $H_{part}$ contains within it a free particle piece which
could serve
as a starting point for an approximation. It is given by the  $i=j$
part of the  dipole-dipole term
\begin{eqnarray}
H_{part}^{free} &=& {l_{0}^{4  }} \sum_{q}^{Q}  \sum_{j} {\pi e^2\over
q}(q^2
|\Pi_j|^2\sin^2 \theta_{q,\Pi} )\\
&=& \sum_j {|\Pi_j|2 \over 2m^*}\\
{ 1\over m^*} &=&  l_{0}^{4} \sum_{q}^{Q} {\pi e^2 \over q}q^2 \sin^2
\theta_{q,\Pi} = {e^2l_0 \over 6}(2\nu )^{3/2}
\end{eqnarray}
where we have used the fact that $Q^2 = 4\pi n = eB \cos^2 \mu
\lambda_0$ and
$\cos^2 \mu \lambda_0 = 2p/(2p+1) = 2\nu$. Notice how the interaction
has
generated  an effective $1/m^*$. Its exact value is not universal and
depends
on the rounded Coulomb interaction employed.

Read has argued\cite{read2} that  the origin
of the kinetic energy is the binding of the CF to the correlation
hole, and that the effective mass is the curvature of this potential
energy{read1}. (The curvature in {\em real } space gives the mass because the
momentum of the fermion is a measure of its dipole moment.)  If we
consider the potential we have used, (the Coulomb interaction cut off
at $Q$ in momentum space, so that it is rounded off to a finite value
at $r=0$), and calculate its curvature at $r=0$, we find that it is
given precisely by the above formula. Of course the bare potential
gives just the bare mass which can be renormalized by the other terms
in $H_{part}$.  It is encouraging that numerical work of Morf and
d'Ambrumenil\cite{morf} gives
$1/m^* = .2e^2 l_0$ for the untruncated Coulomb interaction at  $\nu = 1/2$
compared to our $e^2
l_0/6$ for the truncated Coulomb.  In any event, the aim of the above
calculation is  merely to show that we have a means of computing
$1/m^*$ in a zeroth order approximation and that it has the right size
and physical origin. To our knowledge our earlier work \cite{prl} was
the first instance this was done. The exact value for a given
potential is nonuniversal and dependent on small and large $q$'s and
beyond the reach of our theory.

{\em Note that the kinetic energy that comes from the $i=j$ term does not
constitute the electron's self-energy, but rather the interaction energy
between the electron and its correlation hole produced by the  (deficit of)
other electrons.}  One may ask why we did not banish the $i=j$ term of the
Coulomb interaction in the MR from the outset.   The problem is that we only
know how to canonically transform the full charge density, summed over all
particles. This  density  obeys the magnetic algebra and  functions of it  are
gauge invariant. So our strategy  has been to add the $i=j$ term  in the MR,
{\em where it constitutes an innocuous  chemical potential shift}, work with
the product of the full densities in constructing the interaction hamiltonian
and then transform the latter. The magnitude of the kinetic energy so generated
fits our expectations based on Read's arguments and points to   the soundness
of the procedure.

Returning to general fractions, we have to understand why the CF are
weakly interacting and may be described approximately by
$H^{part}_{free}$, for this is essential to  the success of the CF
picture of Jain\cite{jain-cf}. The smallness of $e^2$ does not matter since it
is a
prefactor to the whole hamiltonian.  Looking at the neglected terms,
we see two possible small dimensionless parameters: $1/p$ and $ql_0$.
Our preliminary analysis of matrix elements (relying on the work of
work of Dai {\em et al}\cite{Dai} and Chen {\em et al}\cite{chen})
shows that there is huge window (which gets larger as $\nu \to 1/2$)
where the dipole can dominate the monopole because
$ql^{*}_{0}\sqrt{p}$ is large even though $ql_0$ is small,  $l^{*}_{0}$ being
the magnetic length appropriate to the reduced  field $A^{*}$.  Not having
completed the  detailed  calculation we merely outline a possible
strategy. Suppose we want $K_{00} (q)$, the retarded
$\rho^{old} - \rho^{old}$ correlation function. Since $\rho^{old}$
does not link physical to unphysical states, its two point function in
a gauge invariant ground state can be calculated without regard to the
constraint on intermediate states. However we do not have access to
this gauge invariant ground state. We must argue that the numerical
value of the correlation will not be too different in an approximation
to the ground state. For the latter we may use the $p$-filled Landau
level ground state of $H^{free}_{part}$. We could also include the
neglected interactions terms in Eqn.(\ref{partham}) in an RPA
calculation.  We are in the process of evaluating this strategy by computing
some correlations that are known numerically or experimentally.
\subsection{$\nu =1/2$}

We now focus on $\nu =1/2$ to which a great deal of attention has been given
following the work of HLR.
 They took  the notion of a Fermi surface at $\nu =1/2$ seriously and deduced
many   experimental consequences which     have been  verified\cite{willett}.

HLR calculated numerous response functions in the RPA  working in the CS
representation we began with (before the introduction of the oscillator
coordinates). Let us focus on the following results\cite{hlr}:
\begin{eqnarray}
\int_{0}^{\infty} Im K_{00} \ \omega \  d\omega &\simeq & q^4\\
\int_{0}^{\infty} Im K_{00} \  d\omega &\simeq &q^3 \log q\\
\int_{0}^{\infty} Im K_{00} \ \omega^{-1} \ d\omega &\simeq& 1/v(q) = q \ \ \ \
\ (for\ Coulomb)
\end{eqnarray}
where the plasmon contribution has been removed.

What does our formalism predict for this response function?
The effective hamiltonian,  the constraint $\chi (q)$ and $\bar{\rho}$
 are \cite{prl}:
\begin{eqnarray}
H_{part} &= &\! \sum_i {p_{i}^{2} \over 2m^*} + \! {  l_{0}^{4} \over
2}\sum_{i, j\ne
i } \sum_{q}^{ Q} {2\pi e^2 \over q} ({q} \! \times \!p_i)({q}\!
\times \! p_j)
e^{-iq(r_i -r_j)}\label{parham}\\
\chi (q) &=& \sum_i e^{-iqr_i} + {il_{0}^{2} \over 2} \sum_i (q \times p_i)
e^{-iqr_i}=0\! \! \!\label{low-energy}\\
\bar{\rho}&=& - il_{0}^{2}\sum_i (q
\times p_i ) \ e^{-iqr_i}
\end{eqnarray}

To calculate $K_{00}$, we need an approximation scheme. One  possibility we
explored  starts with the $p^2/2m^*$ term in Eqn.(\ref{parham})  and  includes
the other terms and constraint in the RPA. While this appears  reasonable and
familiar, there are some dangers in this case, as will be explained shortly.
The result in the small $q$ limit is ,
\begin{eqnarray}
K_{00} &=& {\omega_0 q^2 \over 4\pi (\omega_{0}^{2}-  (\omega +
i \eta
)^2 )}\nonumber \\
&+& q^2 l_{0}^{2} {m^* \over 2\pi} [ {1\over 2} -x^2 +
\theta (x^2 -1)
{|x| \sqrt{x^2 -1}} + i \theta (1 -x^2) {x  \sqrt{1 -x^2}}] \ \ \ \ \ x =
{\omega \over qv^*} \nonumber\\
&\equiv&
K_{00,osc} + K_{00,CF}\label{ourk}
\end{eqnarray}
 We find that the answer is not changed (in the small $q$ limit) as we change
the constraint from  $\sum_i e^{-iqr_i} + {il_{0}^{2} \over 2} \sum_i (q \times
p_i)
e^{-iqr_i}=0$ to $\sum_i e^{-iqr_i} =0$.

The above formula includes the plasmon contribution as well. Note that the CF
contribution has no structure other than the particle-hole branch cut.  It
follows from the above that
\beq \int_{0}^{\infty} Im K_{CF,00} \omega^{\alpha} d\omega =
k_{F}^{\alpha -1}(m^*)^{-\alpha} q^{3 + \alpha} { \Gamma (1 + \alpha
/2 ) \over 8 \sqrt{\pi } \Gamma ( 5/2 + \alpha /2 )}\ \ \ \ \ \ \ \alpha =
0,\pm 1 \label{oursum}
\eeq

The ratio of the $\alpha =1$ and $\alpha =0$ integrals, which gives
the average frequency of the excitations,  goes as
\beq
 <\omega > \simeq q.
 \eeq
The $q$ dependence of the first two sum rules agrees with that of HLR (up to
logarithms) but the last is quite different: in our case the compressibility
vanishes as $q^2$ and is independent of the interaction, while theirs goes as
$1/v(q) \simeq q$ for the Coulomb case.  In  the last sum rule   very small
$\omega$'s dominate.
In HLR, this region is controlled  by the overdamped mode, which goes  as
$\omega \simeq i q^3 v(q)$.
This mode arises due to magic gauge cancellations in the infrared.

We did not see such a mode or such magic cancellations. This could be due to an
improper treatment of the constraints by our RPA.
 Could it be   that if the constraints are treated  properly,  negative powers
of $q$  appear in the correlation function  and neutralize the $q^2$ coming
from the dipolar nature of charge?

Let us look at our constraints and the gauge transform they generate.
 While we do not know them exactly, there is no doubt that as $q \to 0$, they
reduce to
\beq
\chi (q) = \sum_i e^{-iqr_i} =0.
\eeq
  Their  action on the particle coordinate and momenta are as follows:
\beq
{\bf r}_j \to {\bf r}_j \ \ \ \ \ \ \  {\bf p}_j  \to {\bf p}_j + {\bf q}
e^{-iqr_j}
\eeq
Thus $\chi $ essentially  shifts the momenta of the particles. This means that
the Fermi circle cannot be a ground state--   it can be moved around by a gauge
transformation. {\em This is reminiscent of  Haldane's  finding   that the
particles like to cluster around each other in momentum space, but with no
preferred origin. In our  analysis  this reflects  a subset of the gauge
symmetries  of the effective hamiltonian. }

Now, in any gauge theory one resorts to gauge  fixing by choosing an operator
whose commutator with $\chi $ is non zero and freezing its value, say to zero.
For example, in QED, the constraint $\nabla \cdot {\bf E} =0$ is accompanied by
a gauge fixing condition  $\nabla \cdot {\bf A} =0$. Such gauge fixing kills
all spurious fluctuations. In our case it may   justify our fixing the   Fermi
sea arbitrarily at the origin. On the other hand, in a gauge theory, even after
gauge fixing, there can be genuine gauge invariant fluctuations of arbitrarily
low energy. We did not see them in our treatment, and the question is whether
they would arise in a different approximation and if so, produce inverse powers
of $q$  that lead to a finite  static  compressibility for short range forces
and and one that vanishes as $q$ for the Coulomb case.

Recently
Halperin and Stern \cite{ady} have given an existence proof that such a thing
can  happen, and indeed  happens in a  model that arises within our formalism.
 Consider the noninteracting hamiltonian Eqn.(\ref{nonintham}), drop the
oscillator part and the magnetic moment part, and set  $\nu = 1/2$ so that $\Pi
=p$. Then
\beq
\!\!\!\!\!\!\! H =\!\!  \sum_i {p_{i}^{2}\over
2m} \!
 - \!\! {1 \over 2mn}\! \sum_i\sum_j
\sum_{q}^{Q}
p_{i} e^{-iq (x_i-x_j)} p_{j} \!
\eeq
In the    $Q=0$ limit, the  hamiltonian  is invariant under a uniform shift of
all momenta. For this reason  choose $Q$  very small. The formula for charge is
still dipolar since that does not depend on $Q$. Halperin and Stern  compute
the density-density correlation in this model and show that inverse powers of
$q$ indeed appear due to gauge cancellations and modify render the system
compressible.

The authors do not claim that
 this is the  problem we set out  to solve --  it has explicit $1/m$
dependence, no Coulomb interactions, describes particles that are  bound to fat
flux tubes and do not obey any particular statistics except outside a distance
$1/Q$.  {\em The role of this calculation, as stated above,  is to demonstrate
that  inverse powers of $q$ can and do  appear in  a model with the same gauge
symmetries, at least as $q\to 0$ and that dipolar particles  can be
compressible.} The authors however state that the same conclusion applies to
the model at hand and promise details soon. This result was  also announced  by
D.H. Lee in a revised version of Ref.(\ref{lee}).

Note that even if we regain the HLR formulas for correlations, the physical
picture is now much more satisfactory. We have dipolar fermions and a special
hamiltonian whose symmetries  allow for compressibility despite the dipolar
coupling. This hamiltonian and its symmetries are  not postulated but derived
from the microscopic theory.
We also have oscillators which contribute additively to Hall conductance to
complete the picture. Thus we reconcile many seemingly contradictory properties
of the CF's.

 One can seek    other approximations besides  our RPA and see if   such
cancellations occur. Another possibility is to use Eqn. (\ref{rhonat}) for
charge so that
\beq
\bar{\rho}  = \left[ \sum_j e^{-iqx_j} -  {i\l_{0}^{2}\over 2} \sum_j q\times
p_j \ e^{-iqx_j}\right] .
\eeq
In this case the constraint
\beq
 \chi  = \left[ \sum_j e^{-iqx_j} +  {i\l_{0}^{2}\over 2} \sum_j q\times p_j \
e^{-iqx_j}\right]
\eeq
commutes with the  charge (and hence the hamiltonian)   to leading order and
gauge invariance will be  easier to implement.

In the meantime  we can  ask if there are kinematical regions where   our RPA
analysis may hold independent of how well it treats the constraints. The  first
two moments of $K_{00}$ and intuitive arguments suggest that higher the
frequency, the better the prospects. With this in mind we  compared our results
to the surface acoustic waves  (SAW)  experiments of Willett {\em et  al}
(1993) in the frequency range of $\simeq 3.7-6.1GHz$.

The fractional shift in the velocity of SAW, as explained to us by
Simon,\cite{simonsaw}  is given by
\beq
\delta v_s/v_s  = 3.2 \cdot 10^{-4}\left[  1 - \Re (v(q)\ K_{00}(q,\omega ))
\right]
\eeq
where $v(q) = 2\pi e^2/q$ and $3.2 \cdot 10^{-4}$ is a material dependent
constant referred to as $\alpha^2/2$ in the literature and $\Re$ denotes the
real part. The formula is set up so that $\delta v_s =0$ for a free fermion
system in zero field, for which $\Re  K_{00}(q, \omega )=1/v(q)$.

Using our result for $K_{00}$ we obtain
  \beq
\delta v_s/v_s  = 3.2 \cdot 10^{-4} (1- {2\pi e^2 \over \varepsilon q}
{m^*\over 4\pi} (ql_0)^2).
\eeq
Feeding in
\beq
{1 \over m^*} = {Ce^2l_0\over \varepsilon }.
\eeq
and
\beq
 q = {\omega \over v_s} = {2 \pi \cdot 10^9 \bar{f}\over v_s}
\eeq
where $\bar{f}$ is the frequency in GHz, we get
\beq
10^{4} \cdot \delta v_s/v_s  = 3.2 - {.036 \over C}\bar{f}\label{oursaw}
\eeq
upon using  $v_s = 3000m/s$, and $k_F = 1/l_0 =93 \mu m^{-1}$. Notice that the
dielectric constant cancelled  out.

Extracting from Fig. 29, of Willett's 1997 review,    the points
$$(\bar{f} , 10^4 \cdot  \delta v_s/v_s) =  (3.7, 1.85), ({4.3, 1.66}),  ({5.4
,1.37}),( {6.1, 1.17})$$ we obtained a least square fit
\beq
(\delta v_s/v_s  )\cdot 10^4  \simeq 2.9 - .28 \bar{f}.
\eeq
  Upon comparing to Eqn.(\ref{oursaw}),  the slope  translates to a value of
$C\simeq 0.13$. (For reference, the zeroth order value  for the rounded Coulomb
potential was $C  \simeq 0.17$.) The intercept, given by the  on the material
dependent constant  $3.2 \cdot 10^{-4}${\bf }  is within  $\simeq 10\%$ of the
data.

 We are aware that comparing theory to experiment (rather than  to another
theory) is lot more involved. We do not claim to understand in any depth  why
our RPA  should work in this kinematical region.  But we are intrigued by this
agreement with experiment and present it as such. We propose to study other
response functions in this kinematical region to pursue  this question.

\section{ Conclusions}
 The aim of this paper was to start with the problem of planar electrons of
mass $m$, charge $-e$ in a perpendicular magnetic field $-B$  and  see  how far
we could go towards understanding the FQHE within  some approximation scheme.
We discussed the fractions $\nu = p/(2ps+1)$ focusing on $s=1$, the extension
to other values being very direct.

We began by following the standard procedure of attaching point flux tubes to
go from the electronic to the CS representation.  In the hamiltonian
description this led to the introduction of the nondynamical CS field
$(\nabla \times )^{-1} 2\pi l \rho $ when $l$ quanta were attached. The average
value of this (due to the average charge) cancelled some or all of the external
field $A$ leaving behind an $A^*$ which either vanished for $\nu =1/2$ or $\nu
= 1/(2m+1)$;  or for $\nu = p/2p+1$  was just right to fill $p$ Landau level
{\em a la} Jain.

At this point we  took  the crucial step of  enlarging  the Hilbert space to
include the canonical pair $a(q), P(q)$ which described transverse and
longitudinal vector potentials with $0<q\le Q=k_F$. Thus the number of
additional degrees of freedom equaled the number of electrons $n$. This led to
$n$  constraints on physical states ($a(q) |physical >=0$). We made a further
canonical transformation that got rid of the
nasty field $(\nabla \times )^{-1} 2\pi l \rho $  and changed the constraint to
$qa = 2\pi l \rho$. The hamiltonian in  this middle representation (MR)  did
not contain any dependent  fields. We found that if we dropped the terms
$H_{I}$ and $H_{II}$ we obtained a solvable model of $n$ plasma oscillators
described by $a,P$ and $n$ particles, either free or in a field just right to
fill $p$ Landau Levels. The ground state (product) wavefunction for this case,
duly projected to the physical sector using the constraint, gave us the well
known correlated wavefunctions of Jain and Rezayi-Read, (except for the overall
projection to the LLL). The oscillator part of the product was the Jastrow
factor associated with flux attachment, but it came with the magnitude and
phase of the correlation zeros, as well as the gaussian factors with right
magnetic lengths.

We identified operators that created Laughlin's quasiholes. The corresponding
wavefunctions had the right Gaussian factors for the quasiholes, and the
normalization factor from which one could infer their
statistics\cite{halperin2}.
We constructed the operator that could be identified with the Read order
parameter, but pointed out some limitations of this identification. Our
composite boson and fermion operators implemented the physical picture
extracted  by studying Laughlin's wavefunction-- that electrons like to bind
to $zeroes$ of the wave function. {\em Only the zeros produce a charge deficit
to which the electron is attracted; it is  not attracted to just flux, i.e.,
the phase of the zeroes.}  It appeared as if every nondegenerate ground state
of the interacting theory had a counterpart in the free theory of oscillators
and particles upon projection. All these results were obtained before turning
on any explicit $e-e$ interaction although it was understood that such an
interactions exists in order  to stabilize  the unique ground state.

There were however some problems with the  MR  that did not surface  in the
quest for wavefunctions. The plasmons had the wrong frequency (unless $A^*$
vanished) and the particles had kinetic energy of order $1/m$, when they were
supposed to have none (in the noninteracting case) or  one of order $1/m^*
\simeq e^2l_0$ in the presence of  interactions. To cure  this,   we performed
yet another transformation to the final representation (FR) to decouple the
oscillators and particles. This was done in the small $q$ limit  and within an
RPA applied to the {\em final}  variables\cite{final}. The result was  as
follows.

 First, the constraints were entirely in the particle sector. Denying  $n$
particles in two dimensions  $n$ degrees of freedom was just right to project
out the LLL physics.  The particles and oscillators separated in the
hamiltonian. The particle  energy completely vanished except for the coupling
of a magnetic moment $e/2m$ to the external field, a coupling whose existence
had been predicted  by SSH.  Thus the desired quenching of particle KE,  i.e.,
$1/m^* =0$ was accomplished.   The oscillators ended up with the correct
frequency. The formula for $\rho^{old}$, the charge density of the electrons,
contained three pieces: one linear in the oscillators, one proportional to the
particle density with a charge $1/(2p+1)$ and a dipole piece. Thus we regained
the correct charge of the CF and the dipolar nature of the charge for $\nu
=1/2$ as  anticipated by Read.  If we dropped the oscillator part of the charge
(valid in the oscillator ground state sector)  we obtained a projected charge
$\bar{\rho} $ which obeyed the magnetic translation algebra of LLL charge
density first noted by GMP\cite{GMP} (in the small $q$ limit). This projected
density was gauge invariant i.e.,  did not mix physical and unphysical states,
and could be multiplied freely in the enlarged space. Charge-charge
correlations and Hall conductance were additive over the oscillators and
particles since the formulae for charge and  hamiltonian  were. The oscillators
were found carry the full Hall current  in the clean limit we are considering.

We did not try to obtain improved wavefunctions from  the  FR since the
transformations back to the electronic  wavefunction
    were no longer simple. This simply corresponds to the fact that projection
to the LLL (accomplished by the passage to the FR in our scheme and by the
${\cal P}$ operator by Jain and Rezayi-Read)  does something very complicated
to the wavefunctions we derived in Section IV. Since we went from the MR  to
FR in first quantization, our main results-- quenching of kinetic energy and
the reappearance of an interaction dependent $1/m^*$, the value of $e^*$, the
constraints, additivity of response function over particles and oscillators--
all apply to the bosonic case ($
\nu =1/(2s+1)$) as well. Also all the results  in the FR, except for the
cancellation of $1/m$ in the noninteracting case, were insensitive to the
choice $Q=k_F$.\cite{bert}

We  turned on the Coulomb interaction and found that the particle sector now
contained a kinetic energy term. A zeroth order value for $1/m^*$ was easily
found. It had the right order of magnitude and physical origin.  Our
hamiltonian had a gauge symmetry generated by the constraints. In the infrared
limit, these  transformations were seen to be a translation in momentum space,
reminiscent of the drifting  Fermi sea seen by Haldane. We proposed a
computational scheme that was quite reliable for gapped states. For  $\nu =1/2$
we described an RPA calculation of $K_{00}$, the retarded
$\rho^{old}-\rho^{old}$ correlator. Two truncations of the constraints gave the
same answer. Two of  the $K_{00}$ moments were the same as that of HLR (up to
logarithms) while the compressibility vanished as $q^2$ in our case versus
approaching $1/v(q)$. We pointed out that the compressibility formula could be
altered by   gauge cancellations that we did not find, cancellations that
have been shown to occur in a related model by  Halperin and Stern\cite{ady}.
We emphasized that even if in the one ends up with the HLR correlation
functions, our method has facilitated the  reconciliation  many seemingly
contradictory properties of the composite fermion such as its dipolar charge
and compressibility.

We went on and compared our results to the SAW experiments of Willett {\em et
al} on the expectation that perhaps in that kinematical region the precise
treatment  of constraints  would be  unimportant. We  presented the   agreement
with data as an intriguing fact that needs to be understood.

In summary, we have proposed a way to extract a low energy theory for the FQH
states. Composite bosons and fermions with the frequently quoted  properties
arise at tree level in the FR, but one finds that the oscillators are needed
for a complete description. The most important ingrdients are  explicit
expressions for the hamiltonian and charge-current  operators of the CF and CB.
Our description relies on a small $q$ and RPA  approximation (in the final
variables), the precise nature of  which we do not fully understand. The same
applies to the  gauge constraints or gauge symmetry in the FR, of which we are
sure only of the extreme small $q$ limit.   As  we, along with  others ,
continue to work on  clarifying these  issues, we have provided here  many
 details of our machinery in the hope that they may be fruitfully employed  by
a wider audience.

\subsection*{Acknowledgements}
In writing this long version, we have once again profited from the generosity
of numerous colleagues, especially  S.M. Girvin,   J. Jain,   D.H. Lee, N. Read
and
S. Simon.  We are particularly grateful to Bert Halperin for his extensive and
insightful  remarks and comments.  As for errors that crept in despite  the
above, we will blame each other.  We are  pleased to acknowledge the NSF grants
DMR- 9311949       (GM) and  DMR-   9415796  (RS) .

\section*{Appendix A}
We show here that the short distance terms $H_{\delta a}$ along with
the $i \ne
j$ terms in the third sum in Eqn.(\ref{nonintham}) that we shifted to
large $q$ with
a change of sign, conspire to produce the plasmon pole of correct
location and
residue in the case $\nu =1/2$, when $\Pi = p$. Let us focus on just
this part
of the hamiltonian, again
keeping only
zeroth order terms in $q$. (Note that we do not claim $q$ in this
region is
small, only that we are working consistently to a given order in
$q$.)  Recalling that in this region $\rho^{old} = \rho$,
\begin{eqnarray}
\!\!\!\!\!\!\!\!\!!\!\!\!\!  H_{sr} &=& {1 \over 2mn} \sum_{q=Q}^{\infty}
\sum_i\sum_{j\ne i} p_i
\cdot p_j
\  e^{-iq (r_i
-r_j)}
+  \sum_i \sum_{q =Q}^{\infty} {p_i \over       m}e^{iqr_i}\delta
{\bf
a}(q)+
{n\over 2m}\sum_{q=Q}^{\infty}\delta {\bf a}(q)\delta {\bf a}(-q)
\end{eqnarray}
We now switch to the functional formalism. We perform a
Hubbard-Stratonovich
transformation of the first term using a field ${\bf V} = i
\hat{q}V_L - i
\hat{z}\times \hat{q}V_T$. The lagrangian density is (upon dropping
the $\delta
$ symbols in front of the fields, all of which have $q>Q$)
\begin{eqnarray}
L &=& \sum_{p}\sib (p) (i\partial_o - {p^2\over 2m^*})\psi
(p)\nonumber \\
&+& \sum_p\sum_{q=Q}^{\infty}\sib (p+q) \psi (p) \left[ {(i
\hat{q}\cdot p
)V_L(q)\over m}+{( i \hat{q}\times p) V_T(q) \over m}+ {( i
\hat{q}\times p)
a(q) \over m}  -a_0(q)\right]\nonumber \\
& +&\sum_{q=Q}^{\infty}\left[ -{n \over 2m}a(q)a(-q) + {n \over
2m}(V_{T}^{2}+V_{L}^{2})  + a_0(-q) {q \over 4\pi} a(q)\right]
\end{eqnarray}

It is now straightforward to calculate the inverse RPA matrix
propagator for
the 4-component field $D= (a_0, a,V_L,V_T) \equiv (D_0,D_t,D_L,D_T)$
using
\beq
D^{-1} = D_{0}^{-1} + \Pi^0
\eeq
where $\Pi^0$ is the free-field response function. In particular
\beq
[ \Pi^0 ]_{00}(q,\omega )= - \int_{0}^{2\pi} {d \theta \over
4\pi^2}{k_F q \cos
\theta \over \omega - v^* q \cos \theta +i\eta}
\eeq
while the $TT$ (or $tt$) , $LL$, and $0L/L0$ components differ by the
additional factors of $(k_{F}/m)^{2}\sin^2 \theta$, $(
k_{F}/m)^{2}\cos^2
\theta$ and $\pm i(k_F/m ) \cos \theta $ respectively in the
numerator of the
integrand. We do not list these integrals; they may   be found (for
the
time-ordered instead of retarded functions) in the Appendix of
Reference
(\cite{brad}).  Notice that even though $1/m^*$  (or  $v^*$)
vanishes here, we
keep it in the denominator since the answers can have $m^*$ in the
numerator.
Also since we are working to leading order in $q$, we do not keep the
$q^2/2m^*$ in the energy denominator. These functions are needed only
in the
limit
\beq
x = (\omega /qv^*) \to \infty.
\eeq
The inverse matrix in this limit is
\beq
D^{-1} = \left[ \begin{array}{cccc}
 -{m^* \over 4\pi x^2} & {q \over 4\pi} &-{i m^* k_F \over 4 \pi m x}
& 0 \\
 {q \over 4\pi} & -{n \over m} - {m^* n \over 4x^2 m^2} &  0 & -{m^*
n\over
4x^2m^2} \\
 {i m^* k_F \over 4 \pi m  x} & 0 & {n \over m} - {3m^* n\over 4x^2
m^2} & 0 \\
 0 & -{m^* n\over 4x^2m^2} & 0 & {n \over m} - {m^* n \over 4x^2 m^2}
 \end{array} \right]
\eeq
It is now straightforward to verify that as $x\to \infty$ (which is
the only
region we have here) this propagator has only  a plasmon pole at
$\omega_0$. The density-density correlation, related to $D$ by \beq
K = K^0 - K^0DK^0
\eeq
(where $K^0$ is the noninteracting limit)
has the following matrix element:
\beq
K_{00} =  {\omega_0 q^2 \over 4\pi} {1 \over \omega_{0}^{2} - (\omega
+
i\varepsilon )^2 }
\eeq
which  obeys the f-sum rule
\beq
\int_{0}^{\infty} Im K (\omega ) \omega d \omega = {q^2 n \pi \over
2m}.
\eeq

It is important to note that if we had not shifted the $i\ne j$
pieces from
small to large $q$ the plasma pole would have ended up at $4\pi n
/\sqrt{mm^*}$. This term would have also been unwelcome at small $q$
since the
fermions are to have zero hamiltonian. All in all, the shifting from
small to
large $q$ seems decidedly the right thing to do.


\begin{references}
\bibitem{fradbook} Fradkin suggests  in Ref. (\ref{CS}) Eqn.
(10.9.13), another
possible choice of operator. Adapting it to the present case of
time-independent transformations, (replacing the three-dimensional
integral in
the formula by  the two-dimensional one) we find it collapses to the
previously considered
line integral.
\bibitem{ginote} This are some subtle issues here. Both $\rho^{old}$
and $a$
are gauge invariant, as is any linear combination of them. This can
change  if
we drop the $A$ terms in $\rho^{old}$, to project to the LLL. We are
doing this
LLL projection, and also truncating in $q$ and dropping non-RPA
terms. There is
no guarantee that a combination that does not mix physical and
unphysical
states should exist within this truncation, though fortunately it
does.
Finally, although gauge invariant operators will not mix physical and
unphysical states and can be multiplied freely in testing their
algebra, the
converse is not true-- it can be that $\rho^{preferred}$ mixes
physical and
unphysical states, but miraculously has the right commutator. We have
discarded
this possibility.  This expression for charge also seems to have very special
matrix elements between the $p$ and $p+1$ Landau levels.
\bibitem{nonrpa} The neglelcted  non-RPA terms  need some
clarification. A
non-RPA term may be discarded as oscillating in $\bar{\rho}$ or
$\chi$ but may
combine with another  non-RPA term in the evaluation of a product or
commutator to give a nonoscillating term. An
analogy might help. In one dimension, when we  write the
nonrelativistic
lattice Fermi field in terms of a Dirac field obtained by linearizing
near the
Fermi points, the latter will come with oscillating factors $e^{\pm
ik_Fn}$ at
site $n$, leading to terms that   oscillate as $e^{\pm 2ik_Fn}$
in  the
non-relativistic charge density (along with terms that do not
oscillate). The oscillating terms
may be dropped if we are adding a term proportional to the charge
density  in
the hamiltonian. On the other hand, in the charge-charge interaction,
these
will produce terms that go as $^{\pm 4ik_fn}$ which may have to be
kept when
$k_f =\pi/2$ to take care of umklapp processes. The correct
prescription (say in computing a commutator)  is
therefore to keep all terms till we have computed the operator of
interest and
then throw out a terms.
\bibitem{final}
Note that our RPA is done on the particles in the FR rather than in the MR.
The former are weakly coupled (dipolar or with charge $e^*$) so that one may
expect RPA to work better on them. We dropped $H_{II}$ in the MR as RPA for
pedagogical reasons. The dedicated reader may verify that if we keep it and
transform it ot the FR, it will have no RPA pieces.
\end{references}
\end{document}